\documentclass[preprint,10pt]{elsarticle}

\biboptions{sort&compress}

\usepackage{lineno}

\usepackage[utf8]{inputenc}%
\usepackage[T1]{fontenc}%
\usepackage{amssymb,amsmath,amsthm,amstext,amscd}%
\usepackage[margin=2cm]{geometry}%
\usepackage{graphicx}%
\usepackage[%
labelfont=bf, format=hang, format=plain, margin=0pt,
width=0.8\textwidth]{caption}%
\usepackage[list=false]{subcaption}%

\usepackage{hyperref}%
\usepackage[capitalise]{cleveref}%
\crefname{figure}{Fig.}{Figs.}%
\crefname{appendix}{}{}%
\crefname{equation}{}{}%
\usepackage[normalem]{ulem}%
\usepackage{paralist}%

\usepackage{booktabs}%
\usepackage{lscape}

\usepackage{todonotes}%

\usepackage{bm}%
\usepackage[per-mode=symbol]{siunitx}%
\DeclareSIUnit\Length{L}%
\DeclareSIUnit\Time{T}%
\DeclareSIUnit\liter{l}%
\DeclareSIUnit\molar{M}%

\newcommand{\dd}{\mathrm{d}}%
\newcommand{\defeq}{\mathrel{\mathop:}=}%
\newcommand{\Deffpm}{{\mathbf{D}^{\text{eff}}_{\pm}}}
\newcommand{\Deffp}{{D^{\text{eff}}_{+}}}
\newcommand{\Deffm}{{D^{\text{eff}}_{-}}}
\newcommand{\keff}{{\kappa^{\text{eff}}}}
\newcommand{\tp}{{t_{+}}}
\newcommand{\hphiEDL}{{\hat{\varphi}_{\text{EDL}}}}
\newcommand{\phiEDL}{{\varphi_{\text{EDL}}}}
\newcommand{\phiG}{{\varphi_{\Gamma}}}
\newcommand{\hphiG}{{\hat{\varphi}_{\Gamma}}} 
\newcommand{\bchi}{\boldsymbol{\chi}}
\newcommand{\eye}{\mathbf{I}}
\newcommand{\Upore}{\hat{\mathcal{P}}_{\mathcal{U}}}
\newcommand{\Uinterface}{\hat{\Gamma}_{\mathcal{U}}}
\newcommand{\Ustructure}{\hat{\mathcal{S}}_{\mathcal{U}}}
\newcommand{\cin}{{c_{\text{in}}}}%
\newcommand{\lpor}{{l_{\text{por}}}}%
\newcommand{\Nsam}{N_{\text{sam}}}%
\newcommand{\Ntrain}{N_{\text{train}}}%
\newcommand{\Ntest}{N_{\text{test}}}%
\newcommand{\Pb}{P}%
\newcommand{\unif}{\mathrm{Uniform}}%
\newcommand{\given}{\mid}%

\newcommand{\pa}[1]{\mathrm{Pa}_{#1}}
\newcommand{\param}[2]{#1_{#2 | \pa{#2}}}

\journal{Journal of Computational Physics}  
\makeatletter
\def\ps@pprintTitle{%
 \let\@oddhead\@empty
 \let\@evenhead\@empty
 \def\@oddfoot{}%
 \let\@evenfoot\@oddfoot}
\makeatother
\begin{document}%

\bibliographystyle{elsarticle-num} 

\begin{frontmatter}

\title{GINNs: Graph-Informed Neural Networks  for Multiscale Physics}

\author[first]{Eric J.~Hall\corref{mycorrespondingauthor}\fnref{firstfoot}}%
\cortext[mycorrespondingauthor]{Corresponding authors}
\ead{ehall001@dundee.ac.uk}
\author[second]{S{\o}ren Taverniers\fnref{firstfoot}}%
\author[third]{Markos A.~Katsoulakis}%
\author[second]{Daniel M.~Tartakovsky\corref{mycorrespondingauthor}}%
\ead{tartakovsky@stanford.edu}
\address[first]{Division of Mathematics, University of Dundee, Dundee, DD1 4HN, UK}
\address[second]{Department of Energy Resources Engineering, Stanford University, Stanford, CA 94305, USA}%
\address[third]{Department of Mathematics and Statistics, University of Massachusetts Amherst, Amherst, MA 01003, USA}%
\fntext[firstfoot]{Both authors contributed equally to this work.}


\begin{abstract}
We introduce the concept of a Graph-Informed Neural Network (GINN), a hybrid approach combining deep learning with probabilistic graphical models (PGMs) that acts as a surrogate for physics-based representations of multiscale and multiphysics systems. GINNs address the twin challenges of removing intrinsic computational bottlenecks in physics-based models and generating large data sets for estimating probability distributions of quantities of interest (QoIs) with a high degree of confidence. Both the selection of the complex physics learned by the NN and its supervised learning/prediction are informed by the PGM, which includes the formulation of structured priors for tunable control variables (CVs) to account for their mutual correlations and ensure physically sound CV and QoI distributions. GINNs accelerate the prediction of QoIs essential for simulation-based decision-making where generating sufficient sample data using physics-based models alone is often prohibitively expensive. Using a real-world application grounded in supercapacitor-based energy storage, we describe the construction of GINNs from a Bayesian network-embedded homogenized model for supercapacitor dynamics, and demonstrate their ability to produce kernel density estimates of relevant non-Gaussian, skewed QoIs with tight confidence intervals.
\end{abstract}


\end{frontmatter}


\section{Introduction: decision-making using physics-based models and surrogates}%
\label{sec:intro}%

Modeling and simulation of complex nonlinear multiscale and multiphysics systems requires the inclusion and characterization of uncertainties and errors that enter at various stages of the computational workflow. Typically this requires casting the original \emph{deterministic} physics-based model into a \emph{probabilistic} framework 
where inputs or control variables (CVs) are treated as random variables with probability
distributions derived from available experimental data, manufacturing constraints, design criteria,
expert judgment, and/or other domain knowledge (e.g.,~see \cite{Smith:2013uq}). 
Running the physics-based model with CVs sampled according to these distributions yields
corresponding realizations of the system response as characterized by quantities of interest (QoIs). Analysis of the uncertainty propagation from the CVs to the QoIs informs decision-making, e.g., it informs engineering decisions aimed at improving the quality and reliability of designed products and helps identify potential risks at early stages in the design and manufacturing process.

Quantitatively assessing uncertainty propagation presents a fundamental challenge due to the computational cost of the underlying physics-based model. Even for a low number of CVs and QoIs, uncertainty quantification (UQ) for, e.g., accelerating the simulation-aided design of multiscale systems and data-centric engineering tasks more generally (\cite{LauAdamsGirolami:2018dc}), requires a large number of repeated observations of QoIs to achieve a high degree of confidence in such an analysis. The sampling cost is further exacerbated in real-world applications where distributions on QoIs are typically non-Gaussian, skewed, and/or mutually correlated, and therefore need to be characterized by their full probability density function (PDF) rather than through summary statistics such as mean and variance. The computational cost of nonparametric methods to estimate these densities can become prohibitively high when using a fully-featured physics-based model to compute each sample.

One approach to alleviate the computational burden is to derive a cheaper-to-compute \emph{surrogate} for the physics-based model's response enabling much faster generation of output data and thus overcoming computational bottlenecks. Also known as metamodels, emulators, or response surfaces/hypersurfaces, such data-driven surrogate models are statistical models emulating the system response (e.g.,~\cite{EldredEtAl:2004sm,FrangosMarzoukWilcoxEtAl:2010sm}). Their accuracy and fidelity depends on a number of factors including the amount of physics-based model run data available for ``training'' and how the corresponding inputs are selected in the parameter space; they differ from reduced-order model and model hierarchy surrogates which attempt to capture a simplified or lower-fidelity representation of the physical system. However, all surrogates are unified in their aim: computationally cheaper predictions of the response. The use of surrogates in lieu of physics-based representations paves the way for data-driven UQ including sensitivity studies or model calibration with tight confidence intervals for the estimated metrics. 

A plethora of surrogate modeling techniques have been developed for physics-based modeling and simulation including statistical learning of coarse-grained models (\cite{Taverniers:2015ml,DanielsNemenman:2015aa,Harmandaris2016}), radial basis function networks (\cite{Sen:2015,Sen:2018b}), space mapping (\cite{Sen:2018b}), kriging (\cite{Sen:2015,Sen:2017,Sen:2018,Sen:2018b,MakEtAl:2018sm,Sen:2019,Nassar:2019}), polynomial chaos (\cite{UmZhangKatsoulakisEtAl:2017aa,UmHallEtAl:2019bn,TorreEtAl:2019gf}), and neural networks (NNs) (including early works \cite{LeeKang:1990nn,PsichogiosUngar:1992nn,LagarisLikasFotiadis:1998nn,LagarisLikasPapageorgiou:2000nn} and more recent works that take advantage of modern advances in computing, e.g.,~\cite{Balokas:2018nn,TripathyBilionis:2018uq,ZhuZabarasEtAl:2019pc,RaissiPerdikarisKarniadakis:2019pinns}). Machine learning approaches such as NNs and deep NNs, i.e.,~NNs that contain multiple hidden layers between their input and output layers, have received significant attention in recent years thanks to, in part, the advent of off-the-shelf software like TensorFlow \cite{Tf:2015-whitepaper} and PyTorch \cite{Paszke:2019}, which automate the computation of the training loss function gradient via backpropagation \cite{Goodfellow:2016} (a special case of reverse mode automatic differentiation) and tremendously simplify NN design and implementation.

We establish a framework for constructing domain-aware surrogate models, built via deep learning, to support simulation-based decision-making in complex multiscale systems. A fundamental difference between simulation-based decision-making and other scenarios where deep learning is typically used is that in the former setting the user drives the data generation process. With this insight, we deploy a deep learning approach that incorporates well-known strategies from stochastic and predictive modeling in the following way. First, we embed a probabilistic graphical model (PGM) into the physics-based representation to encode complex dependencies among model variables that arise from domain-specific information and to enable the generation of physically sound distributions. Second, from the embedded PGM we identify computational bottlenecks intrinsic to the underlying physics-based model and replace them with a NN. The graph, i.e.,~PGM, informs (i) the selection of the complex physics that the NN learns, (ii) the supervised learning of the NN, and (iii) the predictions made by the trained NN. These last two features are facilitated through the use of structured priors on CVs that serve as inputs to the NN, as highlighted in \cref{fig:NN}, and that differ from the  typical use of independent input layer nodes. We refer to the resulting hybrid PGM/NN surrogate as a ``Graph-Informed Neural Network'' (GINN) as the supervised learning and predictions are guided by the PGM. The PGM-embedded physics-based model yields a domain-aware surrogate and also lends interpretation to the GINN's predictions.

Related to but different from our GINN approach, two main paradigms have emerged with respect to the use of deep NNs for building surrogates of physics-based models described by partial differential equations (PDEs): physics-informed NNs (PINNs) \cite{Raissi:2019,RaissiPerdikarisKarniadakis:2019pinns,
ZhangLuGuoKarniadakis:2019uq,YangPerdikaris:2019nn,Meng:2020} and ``data-free'' physics-constrained NNs \cite{Sirignano:2018,Berg:2018,ZhuZabarasEtAl:2019pc,SunEtAl:2020sm}. Both approaches drive supervised learning by enforcing physical constraints. While PINNs include both the governing PDE and its initial/boundary conditions in the training loss function, physics-constrained NNs enforce the initial/boundary conditions through a bespoke NN architecture while encoding the PDE in the training loss. In contrast, GINNs use simulation data from a domain-aware model without modifying the training loss function. This facilitates their deployment in complex problems that involve a system of PDEs or differential equations and additional constraints of various types. Overfitting in a GINN is controlled using standard non-intrusive and readily available techniques, such as testing the NN on unseen data (i.e.,~data independent of the training samples) and using $\ell_1$ (lasso regression) or $\ell_2$ (ridge regression) regularization (\cite{HastieEtAl:2015sl}).

Further, while GINNs are informed by a graph (e.g.,~a PGM), they are not simply graph NNs (e.g.,~\cite{Scarselli:2009,Zhou:2019}). A typical application of graph NNs is to use a NN for node classification tasks, i.e.,~deciding how to label nodes of a given graph from available labeled data on the remaining nodes. In our approach, the PGM is used to build a domain-aware physics-based model, thus synthesizing stochastic and multiscale modeling. Then computational bottlenecks in the PGM are identified and replaced by a NN whose supervised learning and prediction are further informed by the PGM, e.g.,~through structured priors on CVs.

Our methods are general and can be applied to a wide range of complex models with intrinsic computational bottlenecks. However, here we
showcase the GINN approach via a real-world application of interest in energy storage. The optimal and robust design of electrical double layer (EDL) supercapacitors for use in long-term energy storage devices critically relies on the multiscale modeling of novel nanoporous metamaterials. The presence of nonlinear multiscale physics in this complex system translates into nontrivial correlations both across and within problem scales and necessitates the use of structured probabilistic models (like PGMs) to describe the dependencies among the model variables in order to maintain physically sound distributions. Using a homogenized model of an EDL supercapacitor (\cite{ZhangTartakovsky:2017np}) 
as a computational testbed, we derive a domain-aware physics-based model following the Bayesian Network (BN) PDE framework developed in \cite{UmHallEtAl:2019bn}. The BN PDE model\footnote{We refer to our derived model as a BN PDE since the principal variables involved in the computational bottlenecks are governed by random PDEs.} for supercapacitor dynamics, formulated in \cref{sec:white-box}, is then used to train a GINN surrogate in \cref{sec:black-box}. The GINN replaces the expensive computation of intermediate variables by learned features in its hidden layers. Hence, the GINN replaces random PDE mappings from CVs to QoIs with a NN surrogate model that is domain-aware and physics-informed. The GINN surrogate can then be leveraged to make sufficiently many predictions to quantify uncertainties with a high degree of statistical confidence, as described in \cref{sec:uncertainty-propogation} for nonparametric kernel density
estimation of QoIs. Finally, \cref{sec:concl} is reserved for conclusions and 
follow-up work.

\section{Domain-aware physics-based models: the BN PDE}%
\label{sec:white-box}%

\subsection{Motivation for the use of structured probabilistic models}

The rigorous homogenization in \cite{ZhangTartakovsky:2017np} enables the derivation of macroscopic quantities from microscale counterparts with clearly defined limits of applicability, in contrast to relying on phenomenological relations. While it may be possible to minimize the computational burden with an appropriately chosen simulation technique, such as a multilevel Monte Carlo method as in \cite{Taverniers:2020}, the homogenization and solution of the corresponding closure equations are an \emph{integral feature} of this multiscale physics-based model and thus the associated computational bottleneck is intrinsic. Moreover, the complicated dependencies among the components, some of which are viewed as CVs for the QoIs, demand specialized tools, such as PGMs, to recast the physics-based model into a probabilistic framework.

\begin{figure}[htb]
  \centering
  \includegraphics[width=0.8\textwidth]{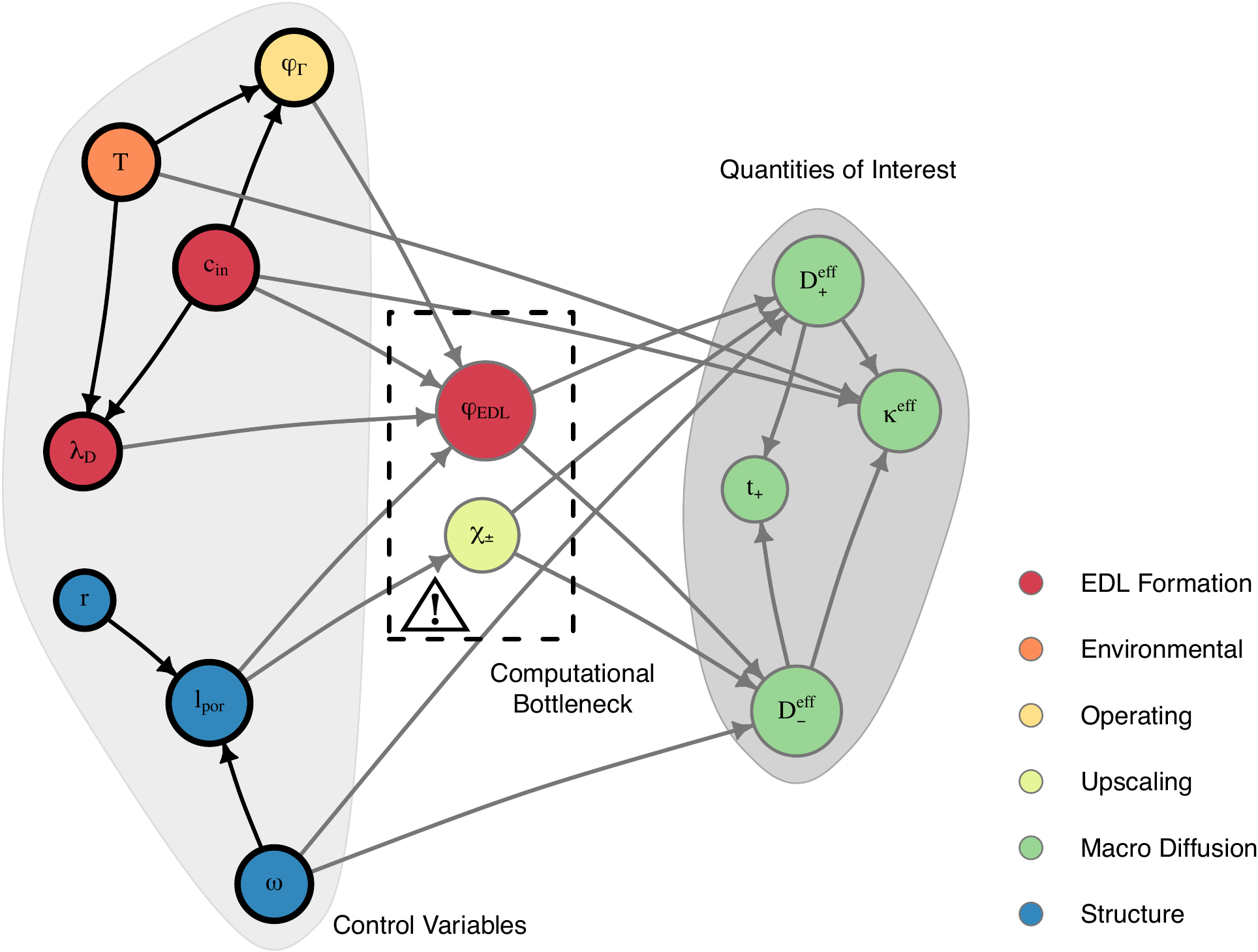}
  \caption{A Bayesian Network (BN), a particular type of PGM, describing supercapacitor dynamics in a nanoporous material encodes conditional relationships for key variables (see  
    \cref{sec:forward-model} and \cref{fig:pore-geom}). The BN enables the systematic and intuitive inclusion of domain knowledge into the stochastic multiscale model and ensures the resulting BN PDE makes physically sound predictions. PGMs guide the supervised learning of GINNs to overcome intrinsic computational bottlenecks in the physics-based model (cf.~see \cref{fig:NN} and \cref{fig:flowchart}).}
  \label{fig:bn}
\end{figure}

Since their introduction, PGMs have proven to be a fundamental mathematical concept for modeling uncertainty in artificial intelligence \cite{Pearl:2014ai,Pearl:2009ci} and machine learning \cite{KollerFriedman:2009gm}. BNs are a class of PGMs that can be represented by a directed acyclic graph with nodes representing random variables and edges representing conditional dependencies. The directed nature of BNs makes describing dependencies intuitive and is therefore well-suited to physics-based modeling. Such structured probabilistic models are necessary in the context of complex systems as, e.g., independent selection of CVs will often lead to non-physical predictions. In \cite{UmHallEtAl:2019bn}, a stochastic modeling framework is presented for embedding BNs into physics-based models. The resulting BN PDEs are random PDEs that incorporate a BN, thereby encoding correlations into stochastic models and providing a platform for uncertainty propagation. More specifically, BN PDEs use the hierarchical structure of BNs to bring together both statistical and multiscale mathematical modeling in a systematic way by informing the physics-based model with domain knowledge including available data, which are typically sparse or incomplete, along with expert opinion, engineering design constraints, and dependencies between CVs. For the problem of interest, a BN encoding the supercapacitor dynamics is shown in \cref{fig:bn}.

\subsection{Formal definition of a BN}

The BN PDE framework centers around constructing a structured probabilistic model for the joint probability density function (PDF) of model variables that captures correlations and constraints among variables in a systematic way. Formally, a BN is defined as a probability model, i.e.,~probability distribution,
\begin{equation}
  \label{eq:BN-def}
  \Pb (\bm{Z} \mid \bm{\theta}) =
  \prod_{i=1}^{n} \Pb (Z_i \mid \pa{Z_i}, \param{\theta}{Z_i})\,,
\end{equation}
for random variables $\bm{Z} = \{Z_1, \dots, Z_n\}$ where $\pa{Z_i}$ is the set of parents of each $Z_i$ and $\bm{\theta} = \{\param{\theta}{Z_i}\}_{i=1,\dots,n}$ are statistical model hyperparameters for each conditional probability distribution (CPD) $\Pb(Z_i \mid \pa{Z_i})$. Here, we assume without loss of generality that the hyperparameters are independent (if not, they can instead be treated as variables).
BNs enable the modeling of large and complex joint distributions containing correlations and the identification of conditionally independent variables significantly reduces the overall dimensionality of \cref{eq:BN-def} thus making parameter inferences from data feasible.

\subsection{Choice of tunable control variables and identification of computational bottlenecks}

We begin by recalling the physical parameters of interest, and their roles, for the dynamics of an EDL supercapacitor described in \cite{ZhangTartakovsky:2017np}; for reproducibility and to provide a self-contained work, the model equations are detailed in \cref{sec:forward-model}. At the macro or continuum scale, the effective ion diffusion coefficients $\Deffp$ and $\Deffm$ (scalar quantities in \cref{eq:effective-diffusion-qois} arising from \cref{eq:effective-diffusion}) are used to compute the effective electrolyte conductivity $\keff$ in \cref{eq:kappa-eff} and transference number $\tp$ in \cref{eq:transference}. Ultimately, these four QoIs are used to inform the state of EDL cells in \cref{eq:edlc-state-model}. The macroscale QoIs depend on microscale parameters, including the solid radius $r$ and pore throat size $\lpor$ of the nanoporous structure (that are also related to the material porosity $\omega$, see \cref{fig:pore-geom}), via a deterministic homogenization (upscaling) with closure $\bchi_{\pm}$ in \cref{eq:closure}. Additionally, the QoIs depend on the temperature $T$, initial ion concentration $\cin$, fluid-solid interface potential $\phiG$ in \cref{eq:transc}, Debye length $\lambda_D$ in \cref{eq:lambda_D}, and EDL potential $\phiEDL$ in \cref{eq:EDL-potential}.

Based on the aforementioned parameters (i.e.,~\cref{eq:effective-diffusion,eq:porosity,eq:closure,eq:EDL-potential,eq:lambda_D,eq:transc,eq:lpor,eq:kappa-eff,eq:transference}), we select thirteen variables,
\begin{equation}
  \label{eq:all-model-vars}
  \bm{Z} \defeq \{\bm{X},\bm{Y}\} = \{X_{\phiG}, X_{\cin}, X_{T}, X_{\omega},
  X_{\lpor}, X_{r}, X_{\phiEDL}, X_{\lambda_D},  X_{\bchi_{\pm}}, 
  Y_{\Deffp}, Y_{\Deffm}, Y_{\keff}, Y_{\tp}\}\,,
\end{equation}
where for simplicity of notation we will use labels instead of indices as in \cref{eq:BN-def} when no confusion arises. The variables
\begin{equation}
  \label{eq:qoi-vars}
  \bm{Y} \defeq \{Y_{\Deffp}, Y_{\Deffm}, Y_{\keff}, Y_{\tp}\}\,,
\end{equation}
represent QoIs that correspond to macroscopic diffusion quantities that parametrize models characterizing the behavior of EDLC cells (cf. \cref{sec:three_eqn_EDL}). The variables
\begin{equation}
  \label{eq:all-input-vars}
  \bm{X} \defeq \{X_{\phiG}, X_{\cin}, X_{T}, X_{\omega},
  X_{\lpor}, X_{r}, X_{\lambda_D},  X_{\bchi_{\pm},  X_{\phiEDL}}\}\,,
\end{equation}
associated, respectively, with electrode surface (fluid-solid interface) potential, initial ion concentration, temperature, porosity, (half) pore throat size, solid radius, Debye length, (upscaling) closure variables, and EDL potential, correspond to both independent and dependent inputs as well as solutions to physical model equations. In particular, the variables
\begin{equation}
  \label{eq:bottleneck}
  \bm{X}_{b} \defeq \{X_{\bchi_{\pm}}, X_{\phiEDL}\} \subset \bm{X}\,,
\end{equation}
represent solutions to random PDEs, i.e.,~the PDEs in \cref{eq:closure,eq:EDL-potential} with random coefficients and/or boundary conditions. These variables correspond to computationally intensive portions of the physics-based model and therefore constitute a computational \emph{bottleneck} for UQ. We investigate the remaining seven variables in \cref{eq:all-input-vars} as tunable CVs,
\begin{equation}
  \label{eq:control-vars}
  \bm{X}_{c} \defeq
  \{X_\phiG, X_{\cin}, X_T, X_\omega, X_{\lpor}, X_r, X_{\lambda_D}\} \subset \bm{X}\,,
\end{equation}
related to the engineering design process and experimental conditions.

We cast the (originally deterministic) homogenized problem into a probabilistic framework by modeling the CVs $\bm{X}_{c}$ as random variables, see e.g.~approach followed for a similar problem in \cite{Taverniers:2020}. The type and support of the distributions placed on $\bm{X}_c$ need to reflect a combination of expert opinion, available data, physical and design constraints, and other domain knowledge, in order to ensure the generation of physically meaningful distributions on the QoIs $\bm{Y}$. While equally valid alternative choices can be made, we select the CVs $X_{T}$, $X_{\cin}$, $X_{r}$, and $X_{\omega}$ to be independent and assume the prior distributions on them to be uniform on an interval of $\pm 35 \%$ (for $X_{T}$ and $X_{\cin}$) or $\pm 25 \%$ (for $X_{r}$ and $X_{\omega}$) around a physically relevant baseline value (see \cref{tab:independent-params}). That is, each of these variables is uniform,
\begin{equation}
  \label{eq:inputs-independent}
  X_{i} \given  \theta_{i}
  \sim \unif([\theta_{i}^{\min}, \theta_{i}^{\max}])\,,
\end{equation}
where the hyperparameters $\theta_i = \{\theta_{i}^{\min}, \theta_{i}^{\max}\}$ represent the minimum and maximum values that are endpoints of the support intervals. 

\begin{table}[htbp] 
  \centering
  \caption{Independent CVs distributed according to \cref{eq:inputs-independent} over the physically relevant ranges, i.e.,~the hyperparameters $\theta^{\min}$ and $\theta^{\max}$, that are selected using expert knowledge and available experimental data (cf.~dependent inputs \cref{eq:inputs-dependent} and in \cref{fig:control-vars-marginal_forward}).}
  \label{tab:independent-params}
  \begin{tabular}{cSSScc}
    \toprule%
    \multicolumn{1}{l}{Variable label} & $\theta^{\min}$ & $\theta^{\max}$ & {Mean/Baseline} & {Variation} & {Units}\\%
    \midrule%
    $T$  & 208 & 432 & 320 & $\pm 35\%$ & \si{\kelvin} \\%
    $\cin$  & 0.52 & 1.08 &  0.80 & $\pm 35\%$ & \si{\mol\per\liter} \\%
    $r$  & 1.05 & 1.75 &  1.40 & $\pm 25\%$ & \si{\nano\meter} \\
    $\omega$  & 0.5025 & 0.8375 & 0.6700 & $\pm 25\%$ & - \\
    \bottomrule
  \end{tabular}
\end{table}

\begin{figure}
  \centering 
    \includegraphics[width=0.45\textwidth]{%
      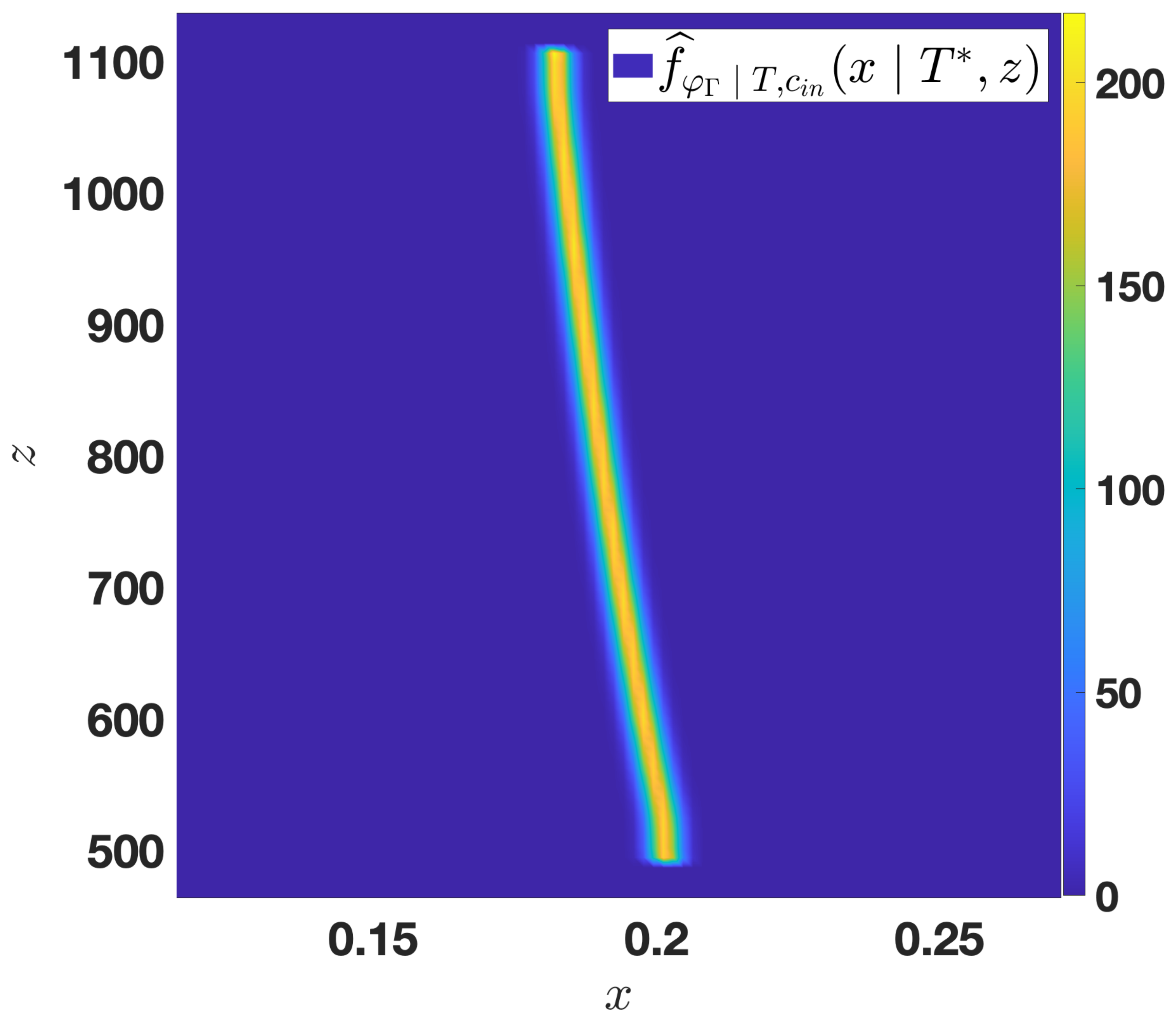}%
    \hfill%
    \includegraphics[width=0.435\textwidth]{%
      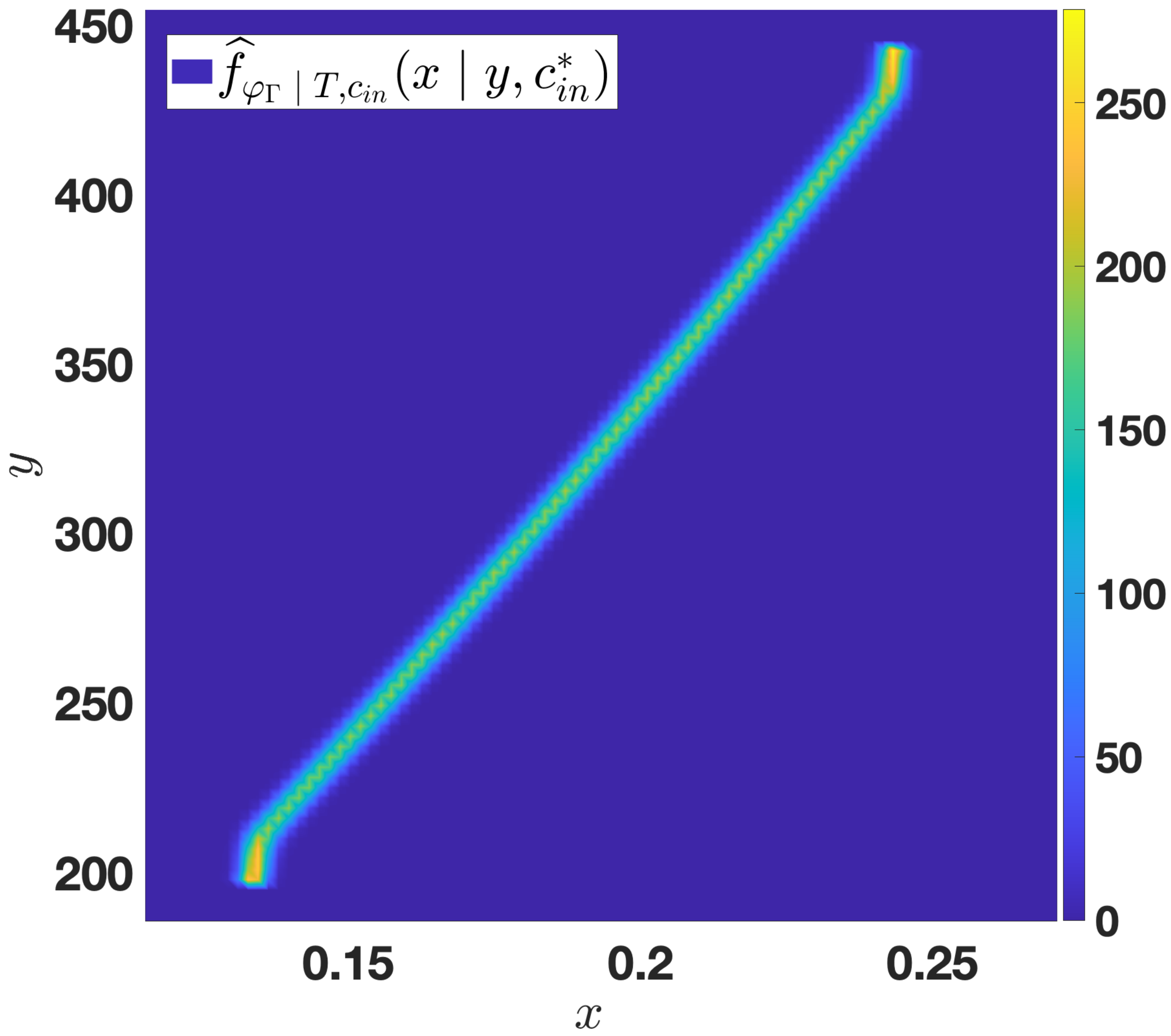}
    \caption{Estimated conditional PDFs, based on $M=\num{1e7}$ samples using kernel density estimation techniques (see \cref{eq:kde_Gaussian_diag}), for the dependent CV $X_\phiG$ in \cref{eq:inputs-dependent:phiG} that is constrained by the nonlinear transcendental equation \cref{eq:phiG-orig} describing its relation to other CVs. From left to right, we show a view of two-dimensional slices of the conditional PDFs, in this case a function of three variables, along $T^*$, the mean value of $X_T$, and along $\cin^*$, the mean value of $X_{\cin}$. The conditional densities \cref{eq:inputs-dependent:lpor,eq::inputs-dependent:lambdaD} can be estimated similarly. }
  \label{fig:control-vars-cpd}
\end{figure}

The distributions of the CVs $X_{\lambda_D}$, $X_{\phiG}$, and $X_{\lpor}$ are then determined by their relation to these independent inputs, captured by \cref{eq:lambda_D}, \cref{eq:transc}, and \cref{eq:lpor}, and the uniform distributions \cref{eq:inputs-independent}. It follows that the conditional distributions on these variables,
\begin{subequations}\label{eq:inputs-dependent}%
  \begin{equation}
    \label{eq::inputs-dependent:lambdaD}
    \Pb(X_{\lambda_D} \given X_{T}, X_{\cin}, \theta_{T}, \theta_{\cin})\,,
  \end{equation}
  \begin{equation}
    \label{eq:inputs-dependent:phiG}
    \Pb(X_{\phiG} \given X_{T}, X_{\cin}, \theta_{T}, \theta_{\cin})\,,
  \end{equation}
  \begin{equation}
    \label{eq:inputs-dependent:lpor}
    \Pb(X_{\lpor} \given X_{\omega}, X_{r}, \theta_{\omega}, \theta_{r})\,,
  \end{equation}
\end{subequations}%
are nontrivial. For example, $X_\phiG$ in \cref{eq:inputs-dependent:phiG} depends on both $X_T$ and $X_\cin$ as the transcendental equation for $\phiG$,
\begin{equation}
  \label{eq:phiG-orig}
    \begin{split}
      \phiG &= \frac{V}{2} - \varphi_\text{ecm} -
    \frac{\sigma}{C_\text{H}}\,,\\
      \sigma &= \sqrt{4\mathcal{E}
        RTz^2 \cin} \sqrt{\cosh\left(\frac{\mathrm{e}\phiG}{k_{\text{B}}T}\right)
        -
        \cosh\left(\frac{\mathrm{e}\varphi_\text{min}}{k_{\text{B}}T}\right)}\,,
    \end{split}
  \end{equation}%
depends on both $T$ and $\cin$ (we refer to \cref{eq:transc} in \ref{sec:forward-model} for a detailed discussion of \cref{eq:phiG-orig}). \cref{fig:control-vars-cpd} displays slices (averages) from the empirical conditional PDF corresponding to \cref{eq:inputs-dependent:phiG} based on $M=\num{1e7}$ observations.
    
These conditional dependencies do not necessarily represent causal relationships. For example, while $\omega$ is an emergent property of $r$ and $\lpor$, we treat $\omega$ and $r$ as independent CVs and which forces us to make $\lpor$ a dependent input (\cref{fig:pore-geom}). Inspired by the recent work in \cite{Li:2020}, this choice allows us to explore numerically a broad range of porosities that was guided by, but not limited to, values in the literature based on prior experiments. In more general terms, such an approach allows parameters for which data are missing over certain ranges to be systematically incorporated and combined with real data through the priors. This being said, investigating the additional inclusion of causal reasoning and its implications for causal inference \cite{Pearl:2016ci} are a natural extension of this line of research.

Finally, we divide the CVs into subgroups associated with environmental conditions $\{X_T\}$, operating conditions $\{X_{\phiG}\}$ (as \cref{eq:transc} depends on the externally applied voltage), structural constraints $\{X_{\lpor}, X_{\omega}, X_{r}\}$, EDL formation $\{X_{\phiEDL}, X_{\lambda_D}, X_{\cin}\}$, and upscaling/homogenization $\{X_{\bchi_{\pm}}\}$. Along with the macroscopic diffusion QoIs $\bm{Y}$, these groupings attach additional layers of significance to the underlying probabilistic model that aid in interpretation and are not necessarily unique. The BN \cref{fig:bn} encodes conditional relationships both between problem scales, via the rigorous pore-to-Darcy scale homogenization (e.g., between macroscopic variables, closure variables, and microscopic structural features), and within single problem scales, such as the geometry and topology of the pore structure. 

\subsection{BN PDE for supercapacitor dynamics}

The joint PDF on all model variables in \cref{eq:all-model-vars} represents the underlying probabilistic model for our application of interest. Using the distributions on the CVs \cref{eq:inputs-independent,eq:inputs-dependent}, which we refer to as structured priors, we decompose the probabilistic model according to \cref{eq:BN-def} to arrive at the BN for supercapacitor dynamics, visualized in \cref{fig:bn}.  This allows us to formally propagate uncertainty from $\bm{X}_c$ via $\bm{X}_b$ to $\bm{Y}$ following the relationships in \cref{fig:bn}. For our application of interest this involves solving a chain of transcendental and algebraic equations, to obtain the dependent CV values, and BN PDEs, associated with the computational bottleneck $\bm{X}_{b}$ in our physics-based model. A visual representation of these steps (to model and propagate uncertainty using the BN and BN PDE) is included in the flowchart for the global GINN algorithm \cref{fig:flowchart}. We shall observe, in \cref{fig:qois-marginal-densities_forward} in \cref{sec:uncertainty-propogation}, that the marginal densities of QoIs $\bm{Y}$ are skewed and non-Gaussian. Moreover, continuous densities are required for downstream computations related to the EDLC cell state model in \cref{eq:edlc-state-model}. Therefore density estimation is the appropriate tool for a corresponding UQ analysis. However, as this requires a large volume of simulations of the physics-based model, in the next section we develop an appropriate surrogate model to accomplish this task.

\section{GINN surrogates for complex systems}%
\label{sec:black-box}%

Predicting macroscopic QoIs with confidence requires repeated solves of the physics-based model. On the one hand, mathematical homogenization is a central feature of the multiscale model that enables the rigorous propagation of uncertainty using the framework in \cref{sec:white-box}. On the other hand, this upscaling is associated with computational bottlenecks that limit our ability to generate sufficiently many realizations of the physics-based model and therefore to carry out a subsequent UQ analysis. Arguably, upscaled models represent the ``best-case scenario'' as direct simulation of the microscale physics is even more computationally demanding in most applications. To address this challenge, we formulate a domain-aware GINN surrogate model suitable for complex systems.

\begin{figure}[t] \centering
  \includegraphics[width=0.8\textwidth]{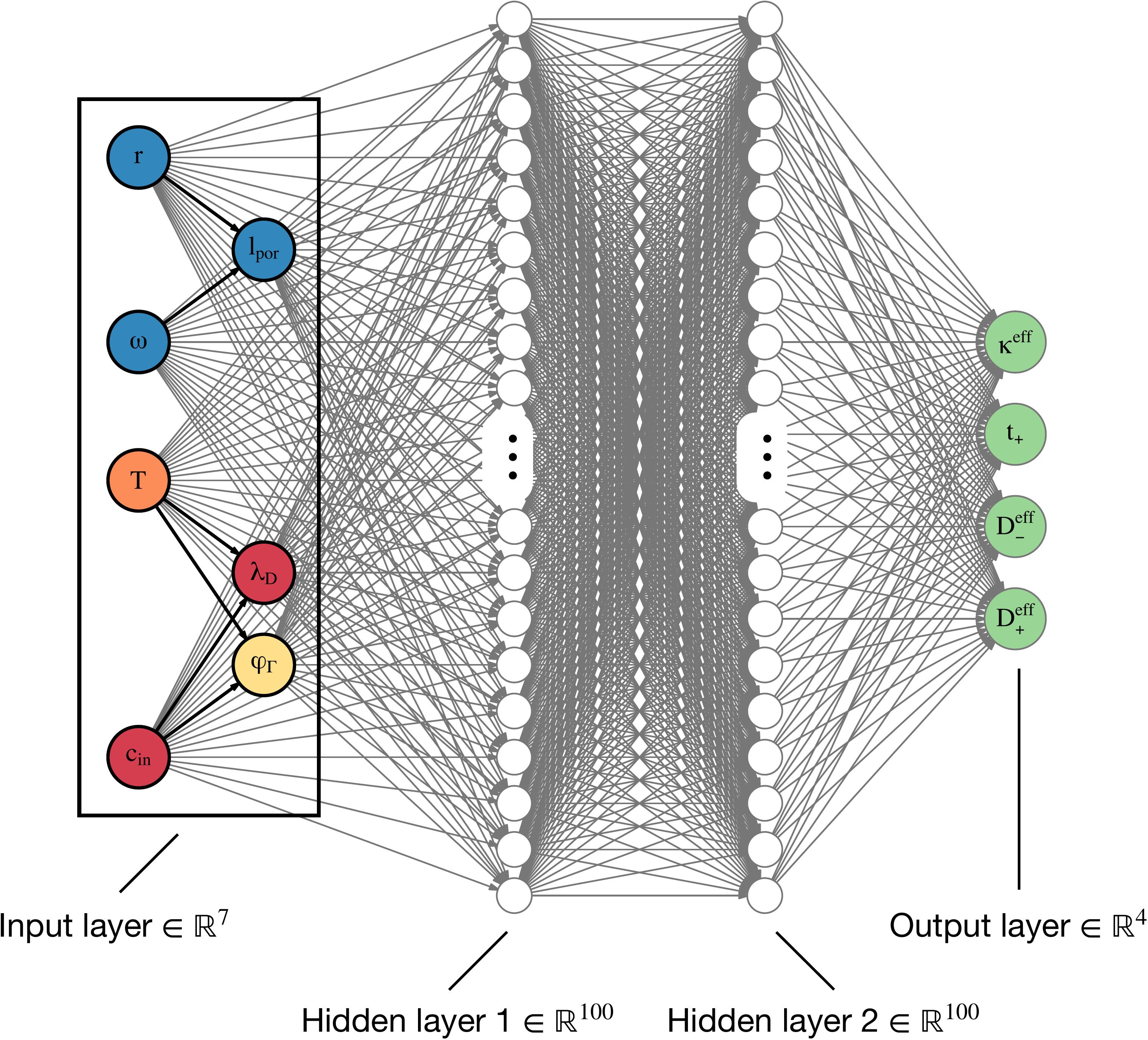}
  \caption{The GINN surrogate for the multiscale model of supercapacitor dynamics. Trained using data simulated with the BN PDE (see \cref{sec:white-box} and \cref{fig:flowchart}), the GINN takes domain-aware graphs as inputs (i.e.,~structured priors on CVs in \cref{fig:bn}) and predicts output  QoIs that bypass computational bottlenecks (highlighted in \cref{fig:bn}) using a deep NN.}
  \label{fig:NN}
\end{figure}

A GINN is essentially a hybrid PGM/NN model that replaces nodes in the PGM associated with computationally expensive solves, e.g.,~corresponding to the homogenization procedure and closure problem, with a deep NN. In the application of interest, rather than solving the computational bottleneck $\bm{X}_b$ \cref{eq:bottleneck} directly using the physics-based model, the GINN learns these intermediate variables as hidden layer features (see \cref{fig:NN}) which, at least conceptually, mirror the boxed nodes of the BN highlighted in \cref{fig:bn}. While replacing nodes in the BN with learned features removes the possibility of making interventions on them, importantly the CVs $\bm{X}_c$ serve as inputs to the GINN rather than being learned features and therefore can still be tuned. Notably, as expressed through the structured priors of the graphical model in \cref{fig:bn}, some of these CVs are mutually correlated, while typically the input layer nodes of NNs are independent. Failure to account for these dependencies would lead to nonphysical distributions for the CVs and hence also for the QoIs.

In lieu of replacing $X_\phiEDL$ and $X_{\bchi_{\pm}}$ in \cref{fig:bn} with a single NN surrogate, our PGM-based representation also accommodates the use of separate PINNs for each of these computational bottlenecks. Likewise, other deep learning approaches, e.g.,~\cite{ZhuZabaras:2018nn,ZhuZabarasEtAl:2019pc,KarumuriTripathyBilionisPanchal:2020sf}, might also be integrated if appropriate for problems where the stochastic dimension is high. While recent work on benchmark problems showed that PINNs can be successfully incorporated into a multifidelity framework where they are trained on both high- and low-fidelity data (``multifidelity PINNs'' in \cite{Meng:2020}), it remains an open question whether replacing multiple systems of PDEs with PINNs in our application of interest yields a tractable multifidelity approach. By retaining the domain knowledge and correlations between CVs in the input layer of the NN, GINNs decouple optimizing the training loss function from encoding relevant physics into the NN and therefore can be easily cast into a multifidelity framework such as the one proposed by \cite{Motamed:2019nn}.

A modeling choice was made to have the GINN predict all four QoIs simultaneously (compare the ``flattened'' output layer in \cref{fig:NN} with the nontrivial dependencies in \cref{fig:bn}) rather than through a separate post-processing step, however the original structure could also be retained. Anecdotally, the discrepancy between the learned samples of $Y_\keff$ and $Y_\tp$ and their counterparts computed by post-processing appears typically to be of $\mathcal{O}(10^{-3})$, which is small compared to other errors and uncertainties considered.

\begin{figure}[htbp]
  \centering
  \includegraphics[width=1.0\textwidth]{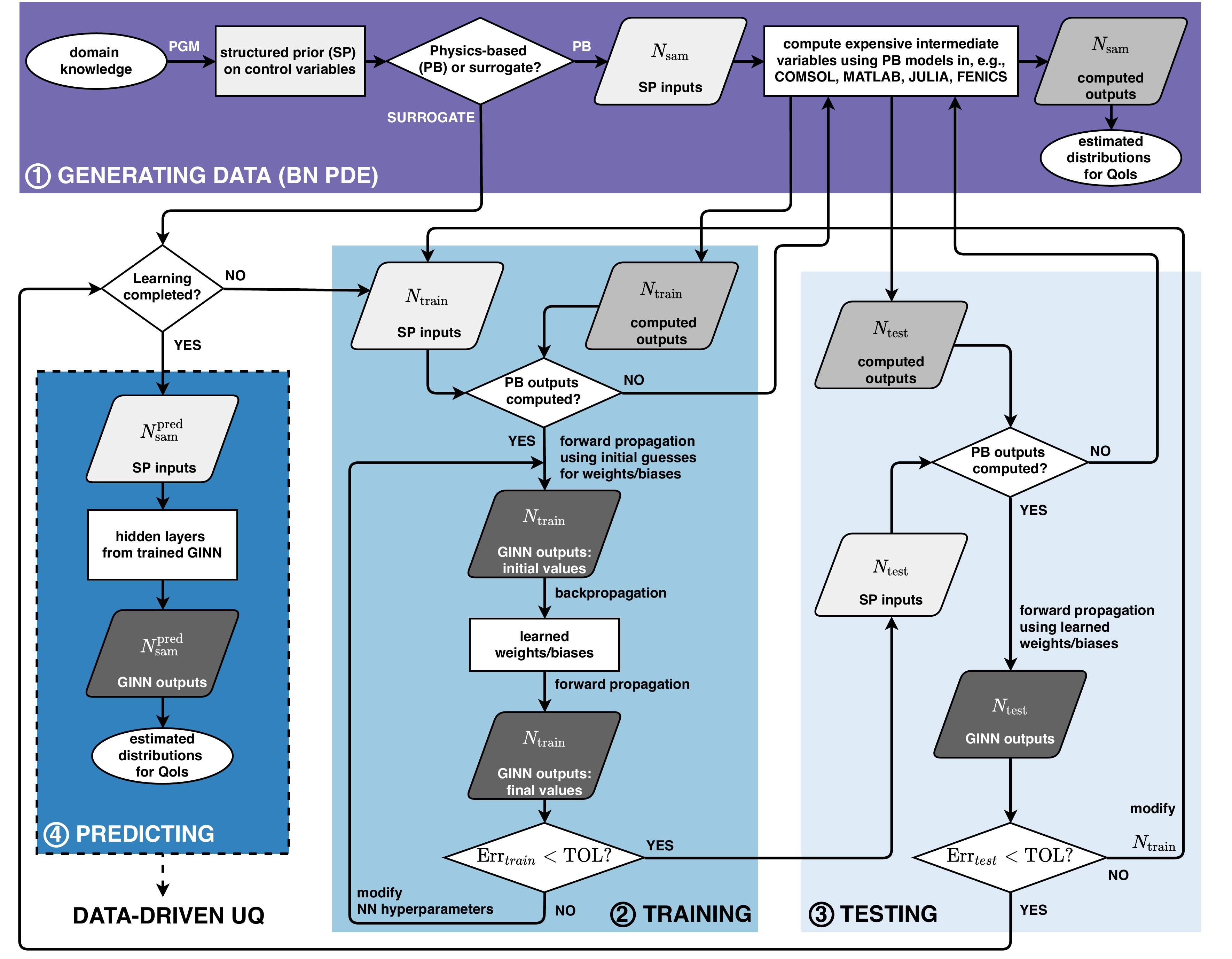}%
  \caption{An overview of the global algorithm for the GINN surrogate. (1) Generation of input-output samples using the BN PDE is expensive because of computationally intensive solves for intermediate variables using physics-based model simulation methods. (2) Training a GINN on a small amount of physics-based model data computed with the BN PDE removes this computational bottleneck by replacing the intermediate variables with learned features (hidden layers). (3) Testing the learned GINN on unseen data avoids overfitting and reduces its generalization error. (4) The resulting trained GINN accelerates the prediction of output data at a much reduced computational cost, thus enabling a rigorous data-driven UQ analysis.}
  \label{fig:flowchart}
\end{figure}

To train the GINN, we obtain simulation data $(\bm{X}_c, \bm{Y})$ comprised of input-output (io) pairs of CVs $\bm{X}_c$ and the resultant QoIs $\bm{Y}$ using the framework in \cref{sec:white-box}. The CVs for our training data are sampled according to the structured prior distributions in \cref{eq:inputs-independent,eq:inputs-dependent} (see also BN in \cref{fig:bn}). These io pairs are then used to train the GINN, depicted in \cref{fig:NN}, using supervised learning. Specifically, the workflow to construct the GINN consists of the following steps which correspond to the numbered boxes in the flowchart for the global GINN algorithm \cref{fig:flowchart}.

\begin{enumerate}
\item \textbf{Generating data (BN PDE):}  We draw $\Nsam = \num{1e4}$ input samples from the structured priors on CVs $\bm{X}_c$ in \cref{eq:control-vars} (see \cref{fig:bn}) and produce $\Nsam$ corresponding samples of the QoIs $\bm{Y}$ in \cref{eq:qoi-vars} using the BN PDE model. These simulations are performed using a co-simulation framework of \textsc{COMSOL} Multiphysics\textsuperscript {\textregistered} and \textsc{MATLAB}\textsuperscript {\textregistered}. This procedure results in $\Nsam$ io pairs $(\bm{X}_c, \bm{Y})$ of simulated data, which we then divide into training and test sets of size $\Ntrain=0.8\Nsam$ and $\Ntest=0.2\Nsam$, respectively.
\item \textbf{Training:} We train the GINN on the training data set through supervised learning. First, we normalize the inputs and outputs to lie on the interval $[-1,1]$ (for the independent CVs) or $[0,1]$ (for the dependent CVs and the outputs) as this increases the rate of decay of the training loss function (mean squared error) as a function of the number of Epochs (one Epoch corresponds to seeing all the training data once). Next, we define a fully connected NN using Google's TensorFlow 2 software (\cite{Tf:2015-whitepaper}) consisting of:
  \begin{compactenum}[(i)]
  \item an input layer comprised of the 7 CVs,
  \item two hidden layers each comprised of 100 neurons, and
  \item an output layer comprised of the 4 QoIs,
  \end{compactenum}
  where all neurons are activated using the ReLU (Rectified Linear Unit) activation function. The structure of the GINN specified above is visualized in \cref{fig:NN}. Finally, we train the GINN on the $\Ntrain$ io pairs using 50 Epochs, which was sufficient to yield a training loss $\text{Err}_{train}$ below a prespecified tolerance $\text{TOL}$ of $\mathcal{O}(10^{-4})$.
\item \textbf{Testing:} This step deals with improving the generalization capability of the model, i.e.,~avoiding overfitting, by validating that it makes suitable predictions on unseen test data. This is done by computing the test loss function (also mean squared error) using the $\Ntest$ test io pairs. If the test loss $\text{Err}_{test}$ is comparable to the training loss, then we proceed to use the trained GINN to predict new io sample pairs.
\item \textbf{Predicting:} We generate $\Nsam^\text{pred}$ inputs for the GINN by sampling from the structured priors, i.e.,~the same distributions as those used to generate the training and test samples. Using the trained GINN, we obtain $\Nsam^\text{pred}$ output samples, each consisting of (normalized) values for the four output quantities, which are then de-normalized to obtain the final predicted QoIs.
\end{enumerate}

Simulation of $\Nsam=\num{1e4}$ io pairs with the BN PDE model using the co-simulation framework (described in \cref{sec:forward-model}) on an Ubuntu system with 8 cores (16 hyperthreads) running at 2.60 GHz and having 64 GB of RAM, given a typical time of 20 seconds per run, takes 3326.4 minutes. Learning the optimal parameter values of the GINN using $\Ntrain=\num{8e3}$ training io pairs and $\Ntest=\num{2e3}$ test io pairs, and generating $\Nsam^\text{pred}=\num{1e7}$ new io pairs using the trained GINN takes about 5 minutes on a 16-inch MacBook Pro (MBP) with 8 cores running at a little under 4 GHz and having 64 GB of RAM. Accounting for the 50\% faster clock speed of the MBP compared to the Ubuntu workstation, predicting $\num{1e7}$ sample pairs with the GINN takes 2222.6 minutes on the MBP (i.e., including the time needed to generate the training/test data and to learn the GINN's parameters). Generation of the same amount of data would take \num{2.2176e6} minutes on the MBP. We conclude that the \emph{cost of generating data with the GINN is almost three orders of magnitude lower than that of generating the data with direct simulation methods}, basically amounting to the ratio between the required number of training/test sample pairs and the number of predicted sample pairs.

Given new values of the CVs sampled according to the structured prior distributions, the learned GINN predicts corresponding samples of the QoIs much faster than would be possible with direct simulation using the physics-based model. This enables the generation of io data sets that are orders of magnitude larger, which drives uncertainty propagation in the next section.

\section{GINN accelerated uncertainty propagation for supercapacitor dynamics}%
\label{sec:uncertainty-propogation}%

By replacing the intrinsic computational bottlenecks of the homogenized model for supercapacitor dynamics by a NN, while retaining domain knowledge and correlations between input nodes, GINNs accelerate the prediction of relevant QoIs and thereby enable data-driven UQ and data-centric engineering approaches to simulation-based decision-making, e.g.,~design of novel metamaterials. A challenge in replacing a well-understood physics-based model with a black box surrogate lies in interpreting and explaining surrogate model predictions. Although we replace the co-simulation framework with a more computationally advantageous surrogate, the complementary BN PDE serves as an anchor for interpreting surrogate predictions. The structured priors on CVs, which encode domain knowledge and constraints, ensure that the inputs to the GINN and hence the resulting output QoIs are physically sound.

Given the possibility of fast generation of sample data using the GINN, we can now estimate marginal and joint densities for QoIs with appropriate confidence intervals. Density estimates form the basic elements for understanding uncertainty propagation between and within scales and are used as building blocks for other UQ tasks such as sensitivity analysis. As the QoIs considered here inform a wide array of phenomena including the three-equation state model for EDLC cells \cref{eq:edlc-state-model}, it is crucial to retain the continuous nature of these variables for downstream computations. Also, as we will demonstrate below, the QoIs for the application of interest are non-Gaussian and skewed. Therefore, density estimates, as opposed to summary statistics, are crucial to provide a complete picture of macroscopic QoIs and their complex interactions.

Kernel density estimation is a nonparametric statistical procedure for estimating probability density functions from samples of a given (univariate) random variable or (multivariate) random vector (see, e.g., \cite{Wasserman:2006np}). For $d$-dimensional identically distributed random vectors $\mathbf{Z}^{(1)}, \dots, \mathbf{Z}^{(m)}$, a kernel density estimator (KDE) for the unknown $d$-variate probability density $f$ is given by
\begin{equation*}
  \label{eq:kde}
  \widehat{f}(\boldsymbol{\zeta}; \mathbf{B}) =
  \frac{1}{M|\mathbf{B}|^{1/2}} \sum_{m=1}^{M} K\left(\mathbf{B}^{-1/2}  (\boldsymbol{\zeta} - \mathbf{Z}^{(m)}) \right)\,,
\end{equation*}
where $K : \mathbb{R}^d \to \mathbb{R}$ is a smooth multivariate kernel function and $\mathbf{B}$ is a $d\times d$ symmetric and positive definite bandwidth matrix. As selection of the kernel shape does not play an important role, we select the widely used Gaussian kernel $K_\text{G}$,
\begin{equation*}
  \label{eq:gaussian-kernel}
  K_\text{G}(\boldsymbol{\zeta}) \defeq \frac{\exp(-\boldsymbol{\zeta}^\top \boldsymbol{\zeta}/2)}{\nu_{d}} \,,
  \quad \nu_{d} \defeq \int e^{-\boldsymbol{\zeta}^\top\boldsymbol{\zeta}/2} \dd{\boldsymbol{\zeta}}=(2\pi)^{d/2}.
\end{equation*}
The choice of the bandwidth matrix $\mathbf{B}$ however is crucial to the performance of $\widehat{f}$, and we choose $\mathbf{B}$ to be diagonal with $B_{ij}=\delta_{ij}b_i^2$, for $i,j=1,\dots,d$, with bandwidths $b_i>0$. 
With these selections of kernel and bandwidth matrix, and defining $\mathbf{b}=(b_1,\dots,b_d)^\top$, \cref{eq:kde} becomes
\begin{equation}
  \label{eq:kde_Gaussian_diag}
  \widehat{f}_Z(\boldsymbol{\zeta}; \mathbf{b}) 
  = \frac{(2\pi)^{-d/2}}{M \prod_{j=1}^d b_j} \sum_{m=1}^M
  \prod_{j=1}^d \exp \left[ -\frac{\left(\zeta_j-Z_j^{(m)}\right)^2}{2b_j^2} \right].
\end{equation}
To automate the computation of bandwidths, we utilize the Improved Sheather--Jones method, a direct plug-in bandwidth selector from \cite{BotevGrotowskiKroese:2010kd}.

As for any estimated quantity, confidence regions can, and should, be given for KDEs. The asymptotic normality of the pointwise error enables one to define a $1-\alpha$ confidence interval pointwise for $\widehat{f}_Z(\boldsymbol{\zeta}; \mathbf{b})$,
\begin{equation}
  \label{eq:kde-ci}
  C_{1-\alpha}(\boldsymbol{\zeta})
  = \left[ \widehat{f}_Z(\boldsymbol{\zeta}; \mathbf{b})
    - z_{\alpha/2} \sqrt{\frac{\mu_{K,d} \widehat{f}_Z(\boldsymbol{\zeta}; \mathbf{b})}{M \prod_{j=1}^d b_j}} \,,
    \widehat{f}_Z(\boldsymbol{\zeta}; \mathbf{b})
    + z_{\alpha/2} \sqrt{ \frac{\mu_{K,d} \widehat{f}_Z(\boldsymbol{\zeta}; \mathbf{b})}{M \prod_{j=1}^d b_j} }
  \right] \,,
\end{equation}
where $z_{\alpha/2}$ is defined through $\Phi(z_{\alpha/2}) = 1-\tfrac{\alpha}{2}$ with $\Phi$ the cumulative distribution function for the standard normal (see, e.g., \cite{Chen:2017ks}) . The parameter $\mu_{K,d}$ is a constant that depends on the kernel $K$ and dimension $d$; e.g., for a Gaussian kernel $K_\text{G}$ and $d=1$, $\mu_{K_\text{G},1} = (2\sqrt{\pi})^{-1}$ (and for $d=2$, $\mu_{K_\text{G},2} = (4\pi)^{-1}$). $C_{1-\alpha}(\boldsymbol{\zeta})$ is an easy to compute random interval such that
\begin{equation*}
  \label{eq:ci-prob}
  \Pb\left( \mathbb{E}[\widehat{f}_Z(\boldsymbol{\zeta}; \mathbf{b})] \in C_{1-\alpha}(\boldsymbol{\zeta})  \right)
  \geq 1-\alpha \,,
\end{equation*}
that is, the confidence interval holds pointwise for $\boldsymbol{\zeta}$.

\begin{figure}
  \centering
  \includegraphics[width=0.495\textwidth]{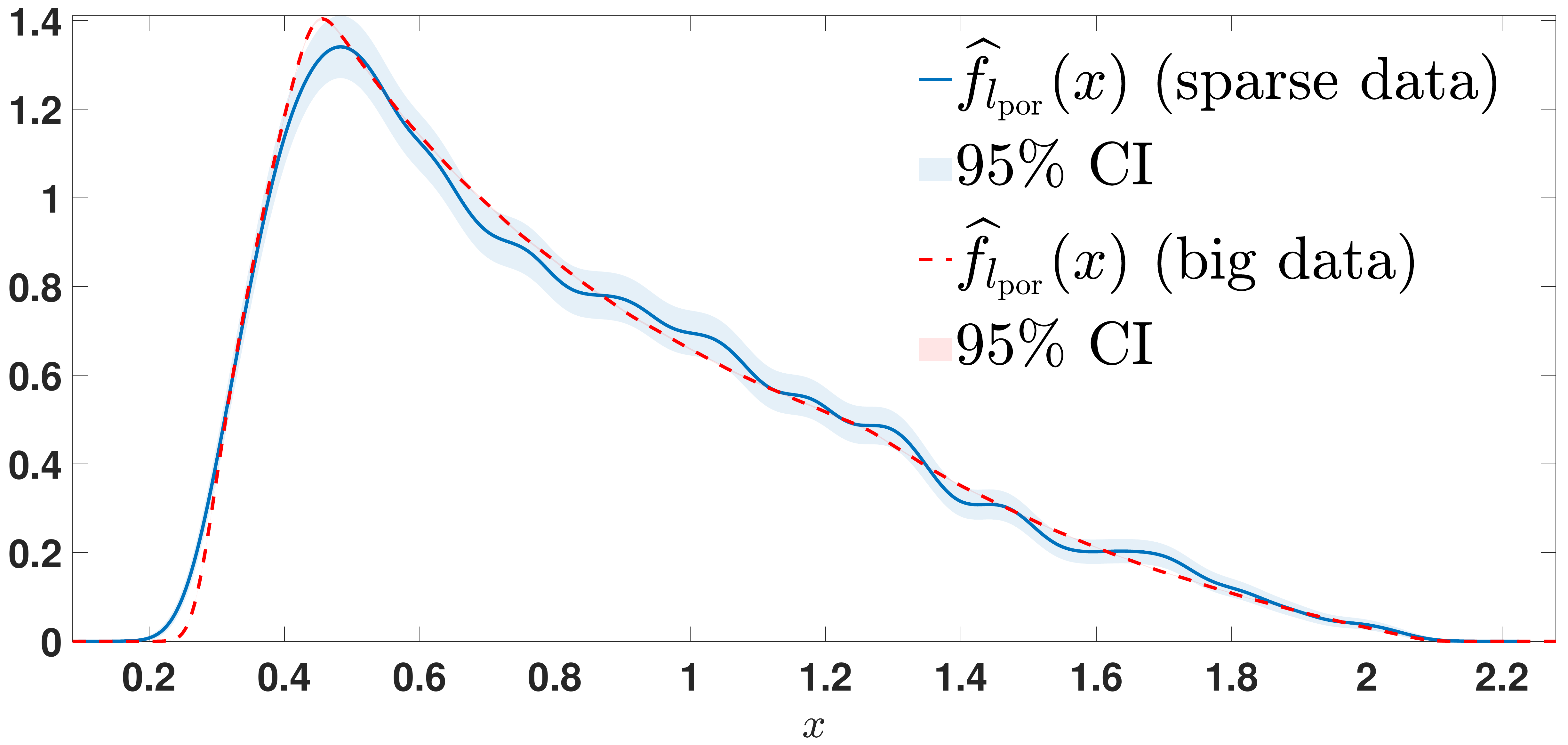}%
  \hfill%
  \includegraphics[width=0.49\textwidth]{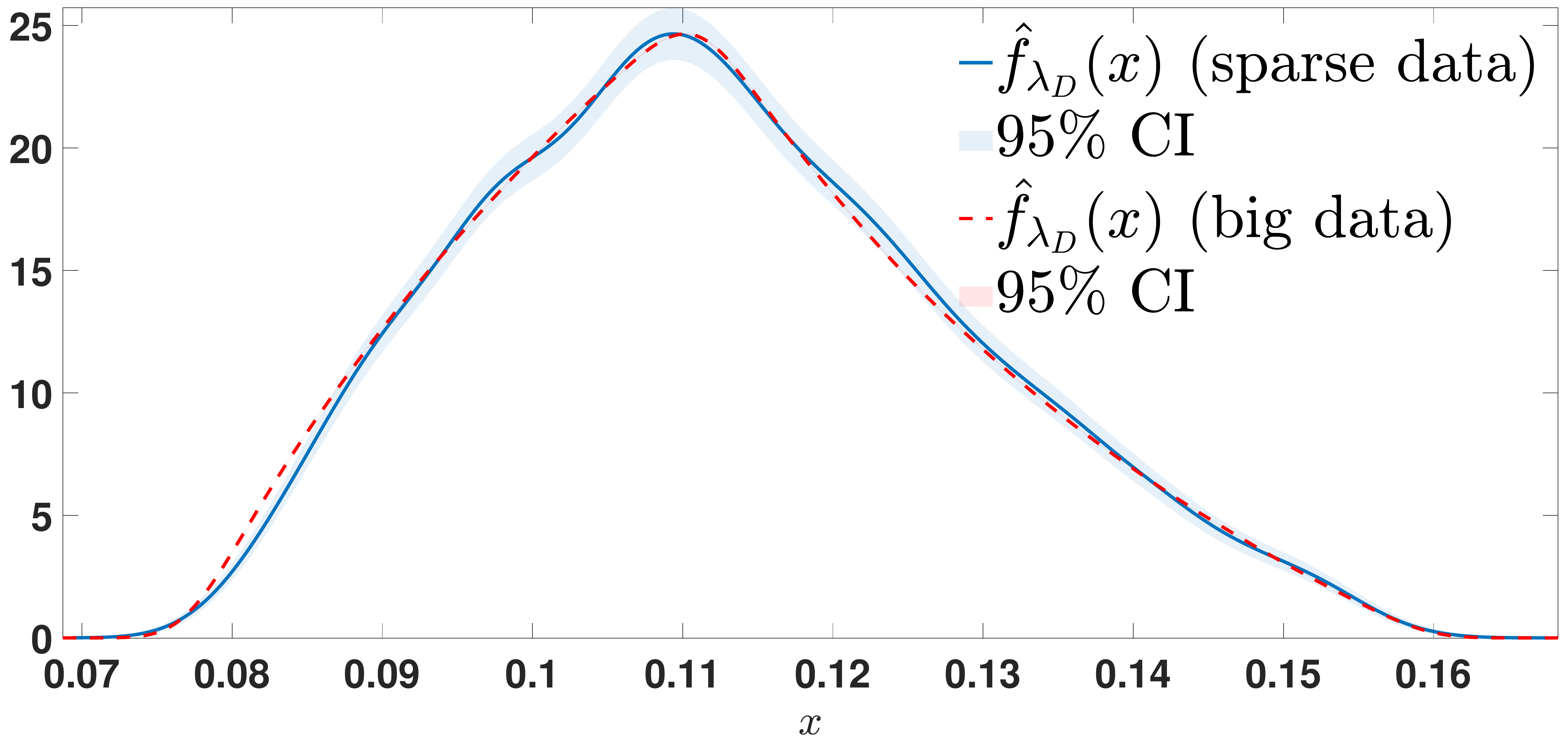}%
  \newline
  \includegraphics[width=0.49\textwidth]{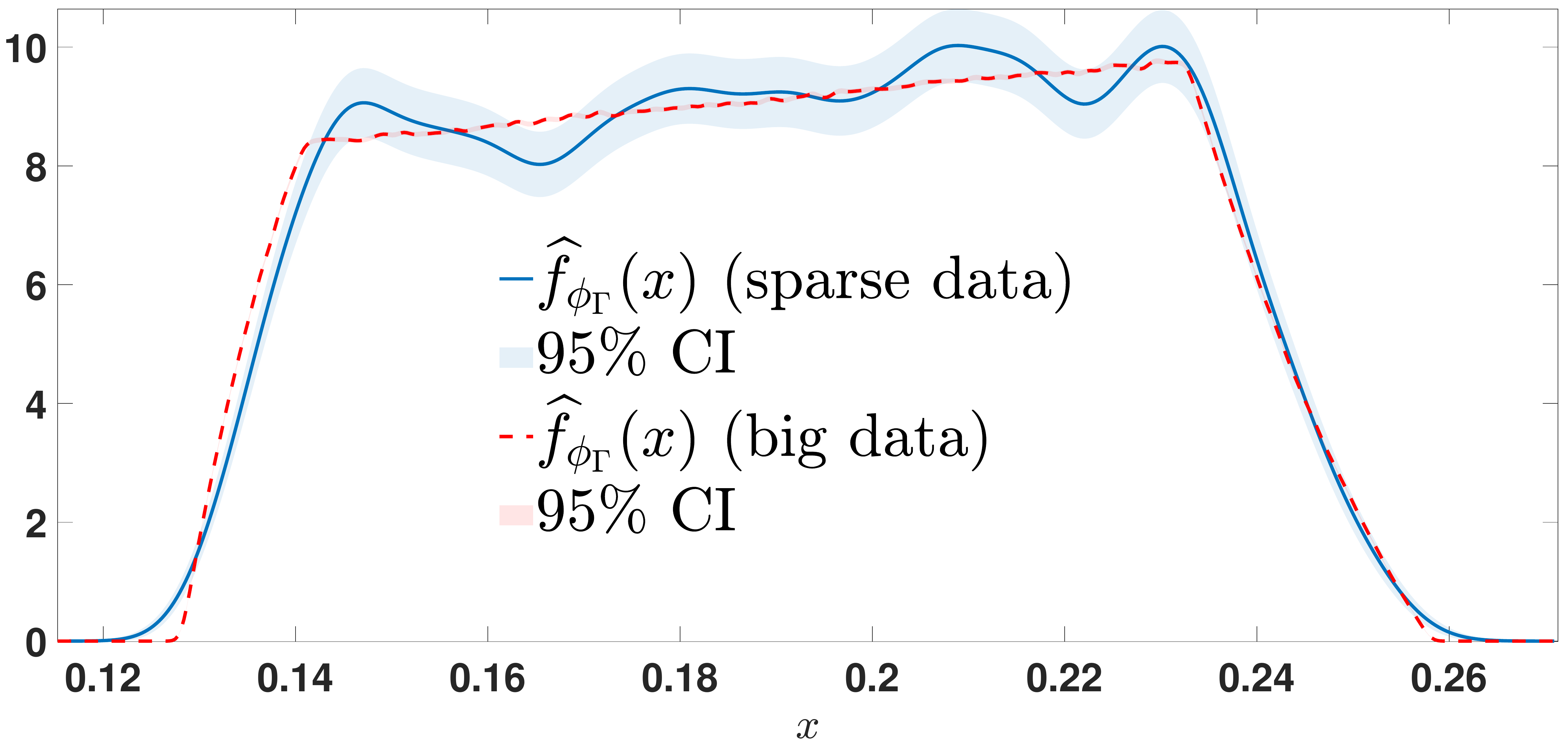}%
  \caption{The reliability of KDEs depends on the amount of data available, as illustrated for the marginal PDFs on dependent CVs in \cref{eq:inputs-dependent} by the spurious features with loose confidence intervals resulting from $M=\num{8e3}$ (solid/blue) observations compared to the smooth densities with tight confidence intervals resulting from $M=\num{1e7}$ (dashed/red).  A large volume of observations is computationally feasible as the the structured priors on CVs, \cref{eq:inputs-independent,eq:inputs-dependent} and  \cref{tab:independent-params}, precede the computational bottleneck in the BN \cref{fig:bn}.}
  \label{fig:control-vars-marginal_forward}
\end{figure}

KDEs for the marginal distributions, \cref{eq:kde_Gaussian_diag} with $d=1$, associated with the CVs $X_\lpor$, $X_{\lambda_D}$, and $X_{\phiG}$ are displayed in \cref{fig:control-vars-marginal_forward} together with appropriate confidence intervals \cref{eq:kde-ci}, based on both $M=\num{8e3}$ and $M=\num{1e7}$ observations. The reliability of the KDEs, i.e.,~absence of spurious features, and the tightness of the confidence intervals very much depend on the amount of available data. As sampling the structured priors for the CVs is relatively inexpensive, the density estimates in \cref{fig:control-vars-marginal_forward} can be obtained with arbitrarily high confidence by making additional observations a priori. In contrast, QoIs, as child nodes of the highlighted bottleneck in \cref{fig:bn}, are expensive to sample using simulation of the physics-based model, such as direct computation of the BN PDE nodes $\bm{X}_b$. Instead, we can use our GINN surrogate model to construct KDEs for the marginal and joint densities of the QoIs with a high degree of statistical confidence.

\begin{figure}
\centering
  \includegraphics[width=0.5\textwidth]{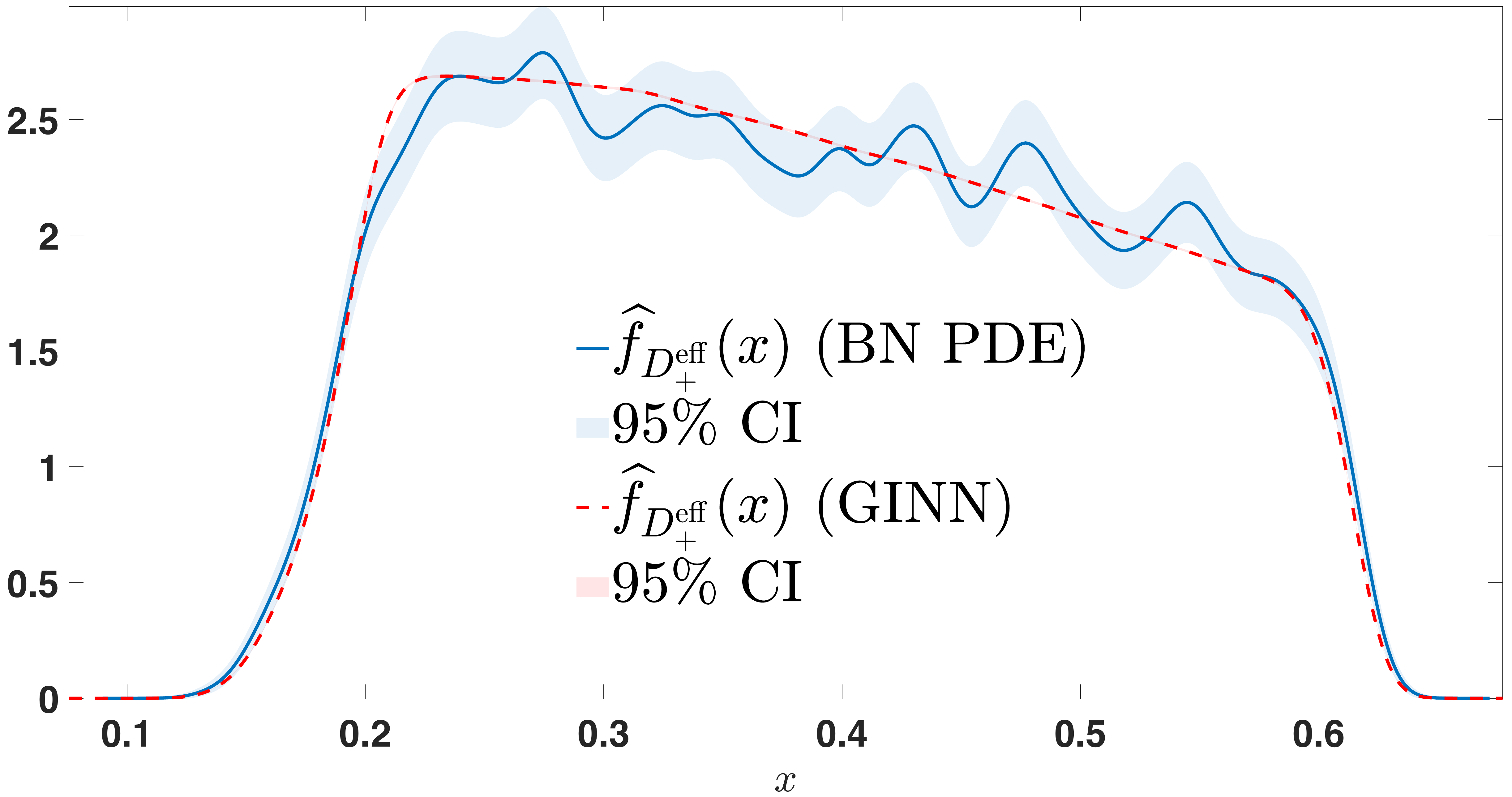}%
  \hfill
  \includegraphics[width=0.49\textwidth]{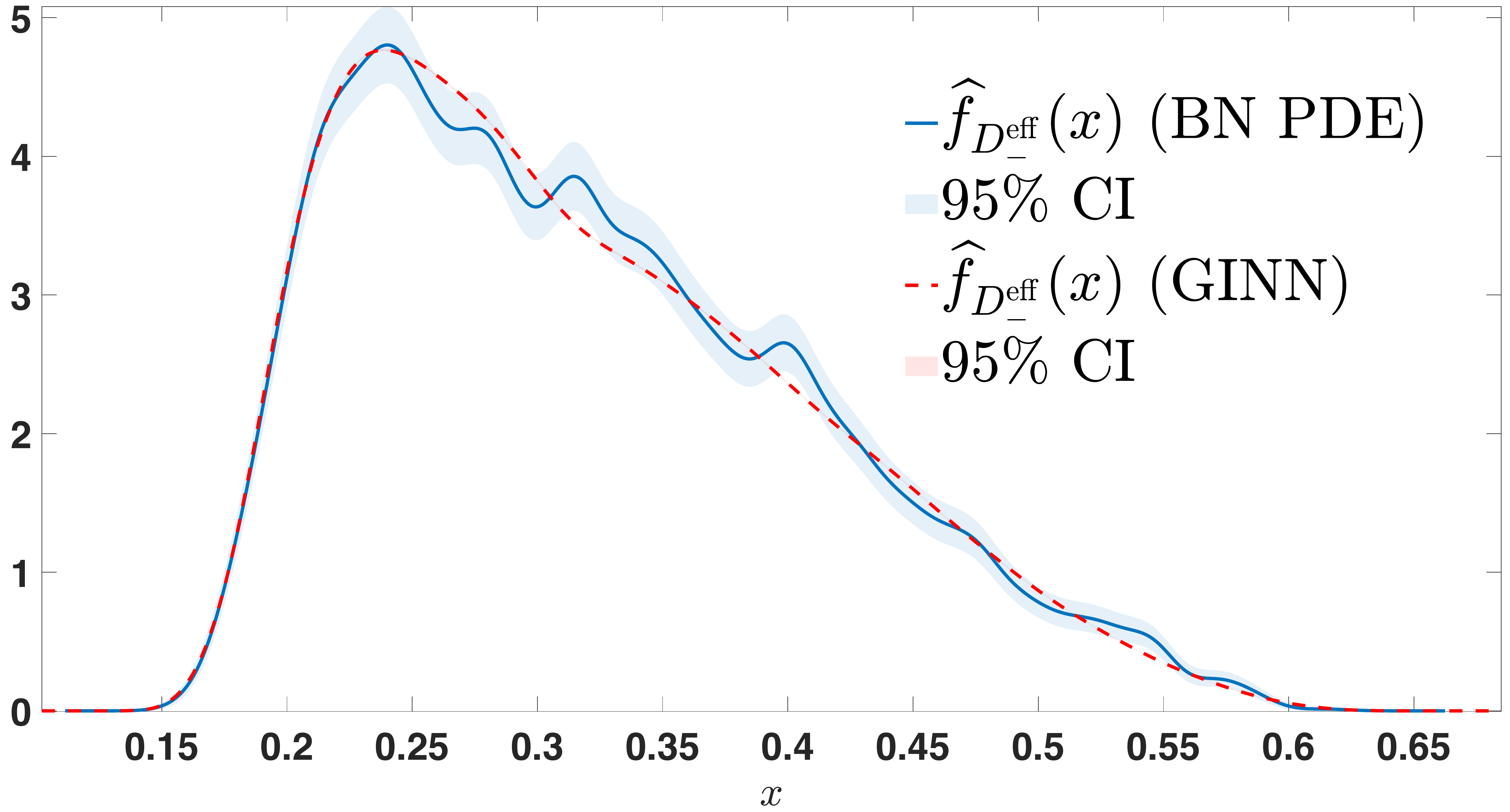}
  \newline
  \includegraphics[width=0.5\textwidth]{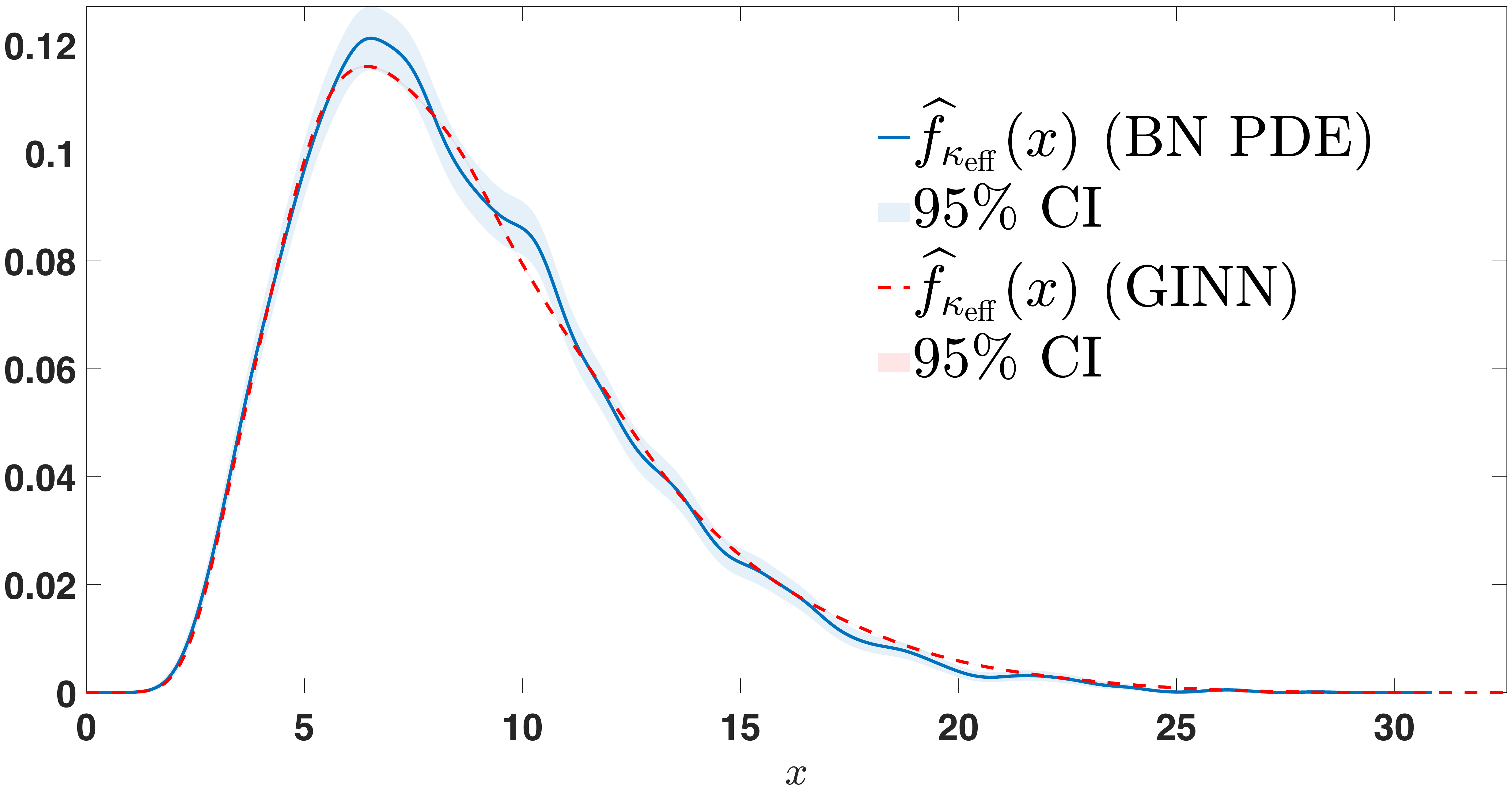}%
  \hfill
  \includegraphics[width=0.49\textwidth]{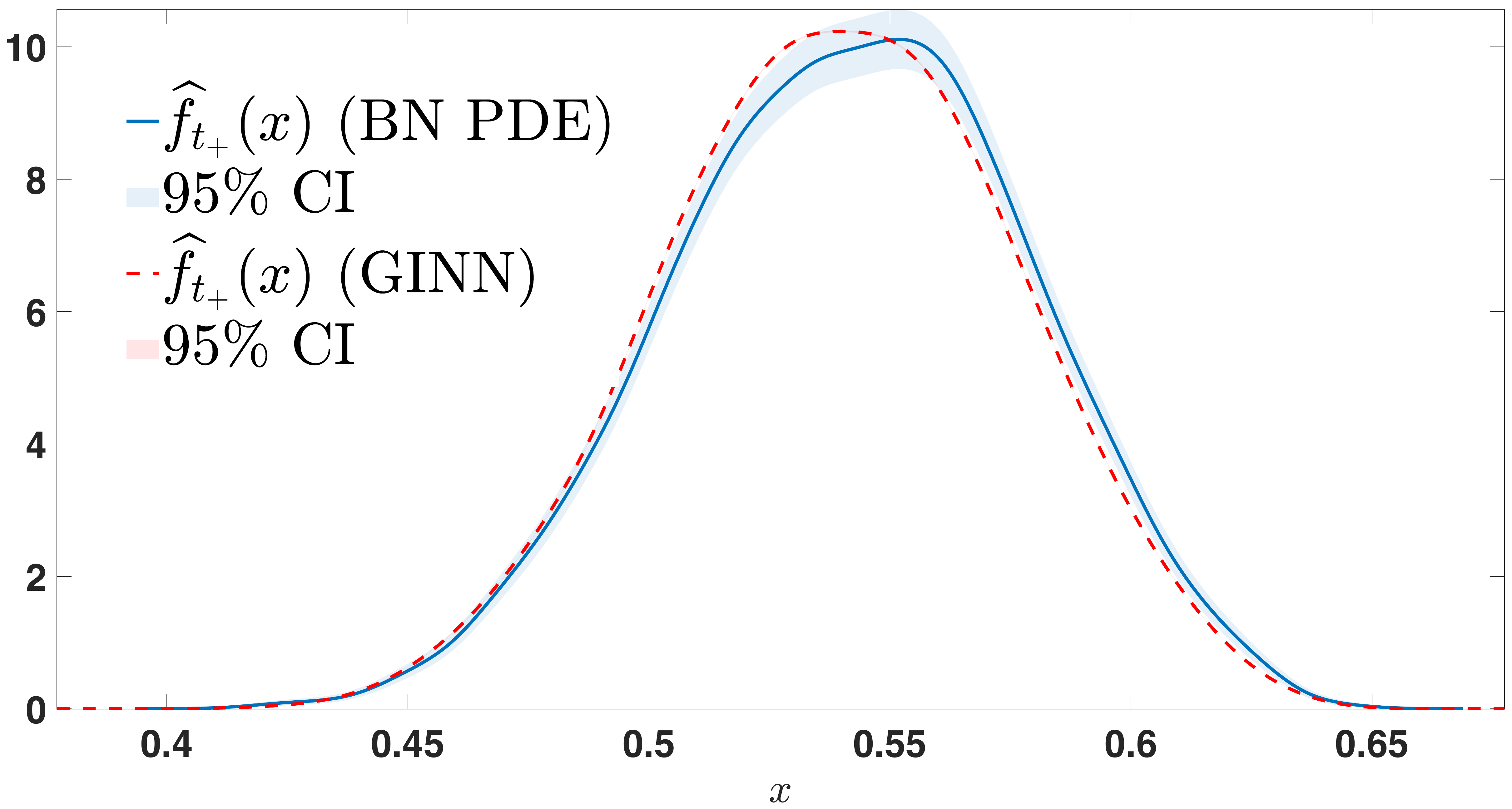}
  \caption{The estimated marginal densities for the QoI variables \cref{eq:qoi-vars}, each based on $M=\num{8e3}$ samples computed with the BN PDE (solid/blue) or $M=\num{1e7}$ samples computed with the GINN (dashed/red), are non-Gaussian and skewed. The higher number of samples that can be generated with the GINN surrogate enables tighter estimates that omit spurious features, thereby enabling data-driven UQ analysis.}
  \label{fig:qois-marginal-densities_forward}
\end{figure}

\begin{figure}
  \includegraphics[width=0.45\textwidth]{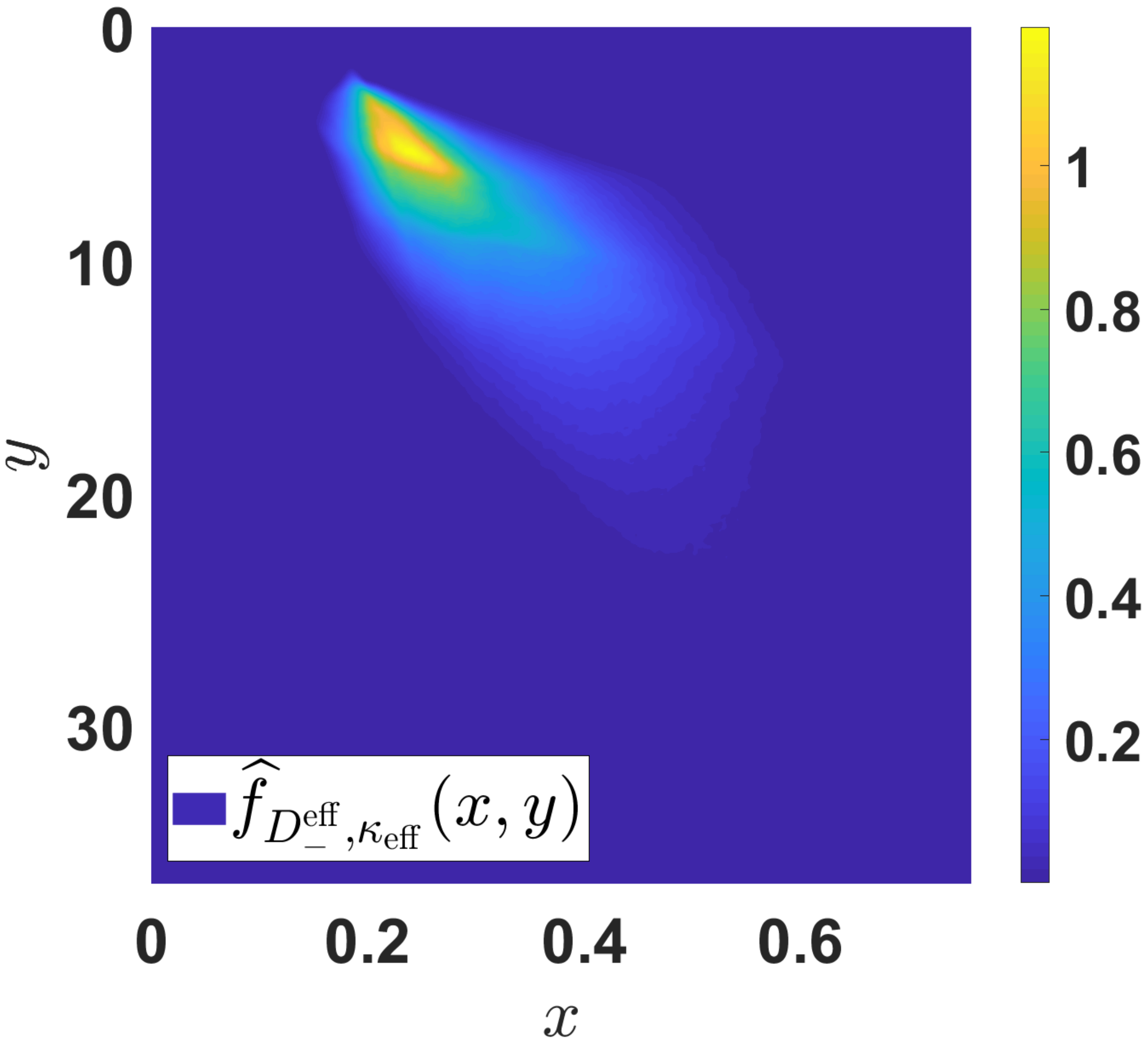}%
  \hfill
  \includegraphics[width=0.45\textwidth]{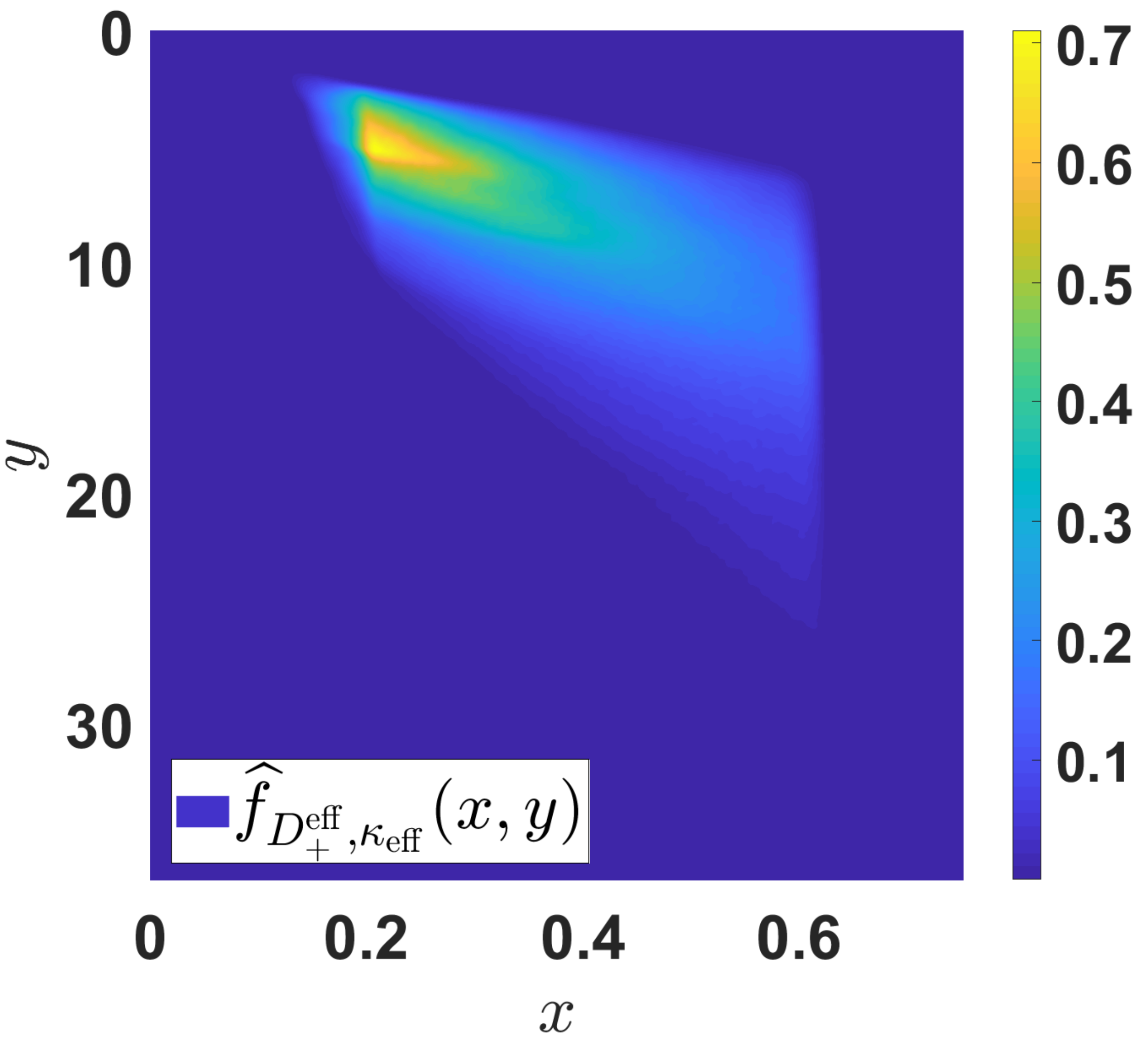}\\
  \includegraphics[width=0.45\textwidth]{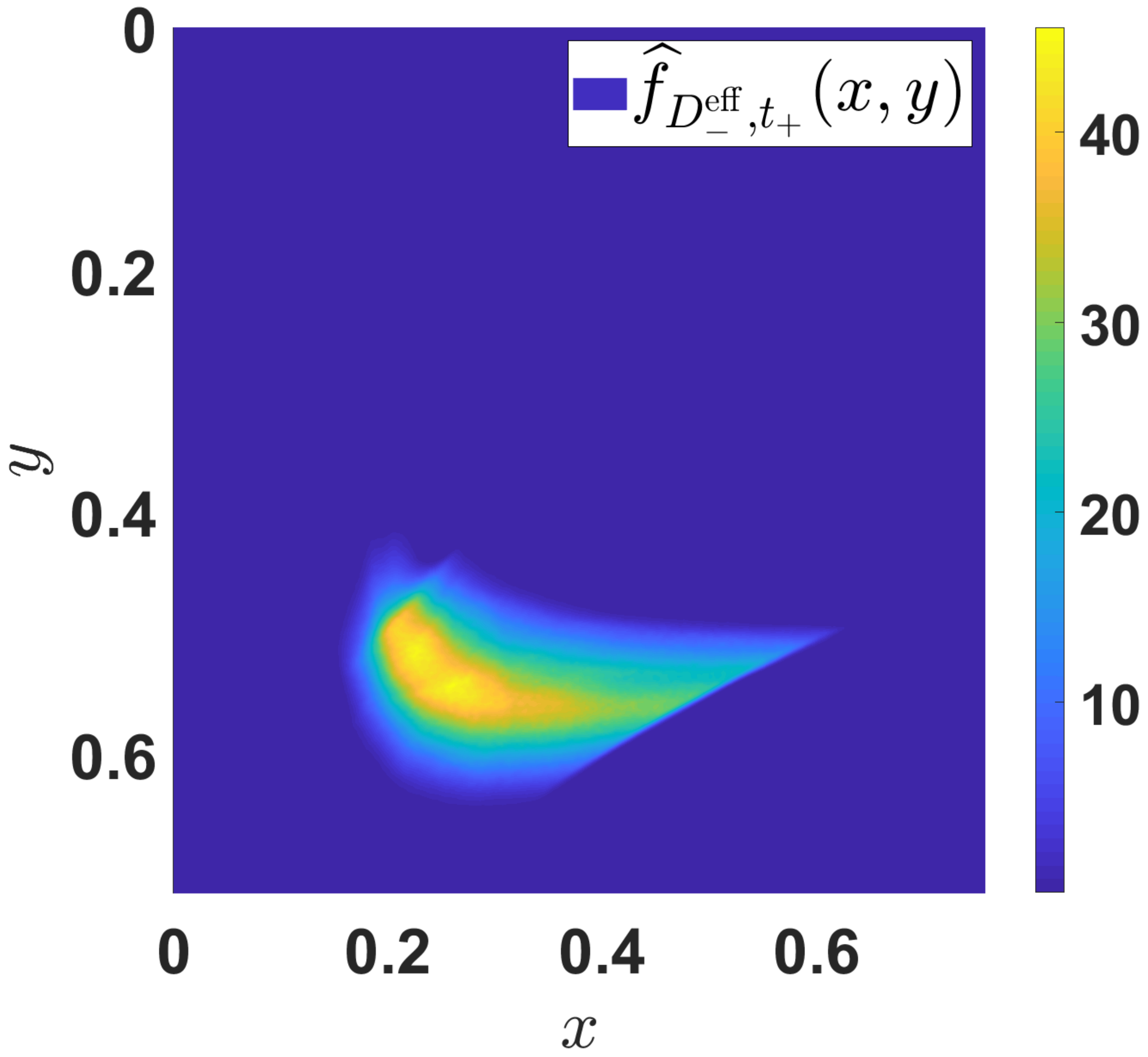}%
  \hfill
  \includegraphics[width=0.45\textwidth]{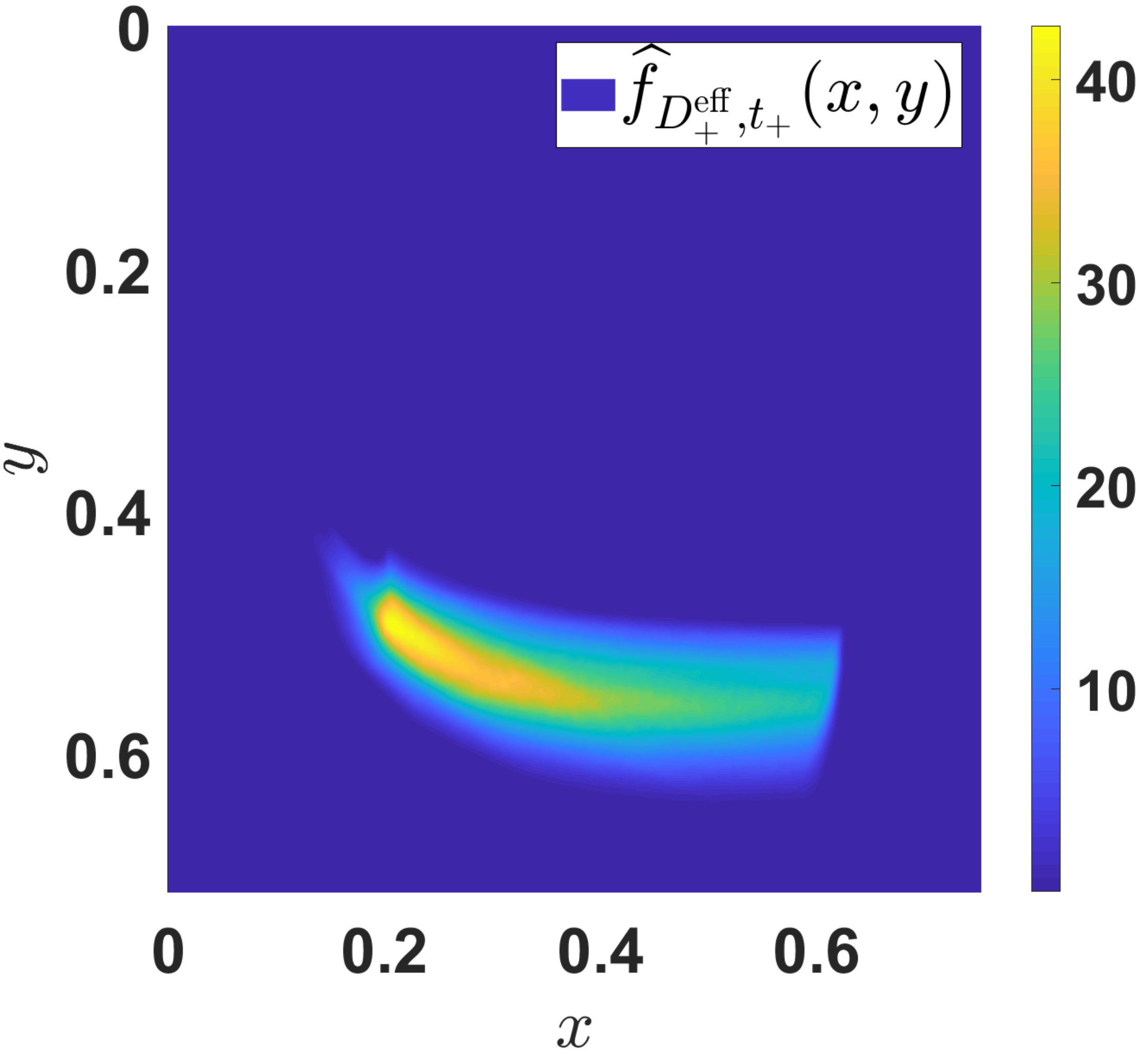}\\
  \includegraphics[width=0.45\textwidth]{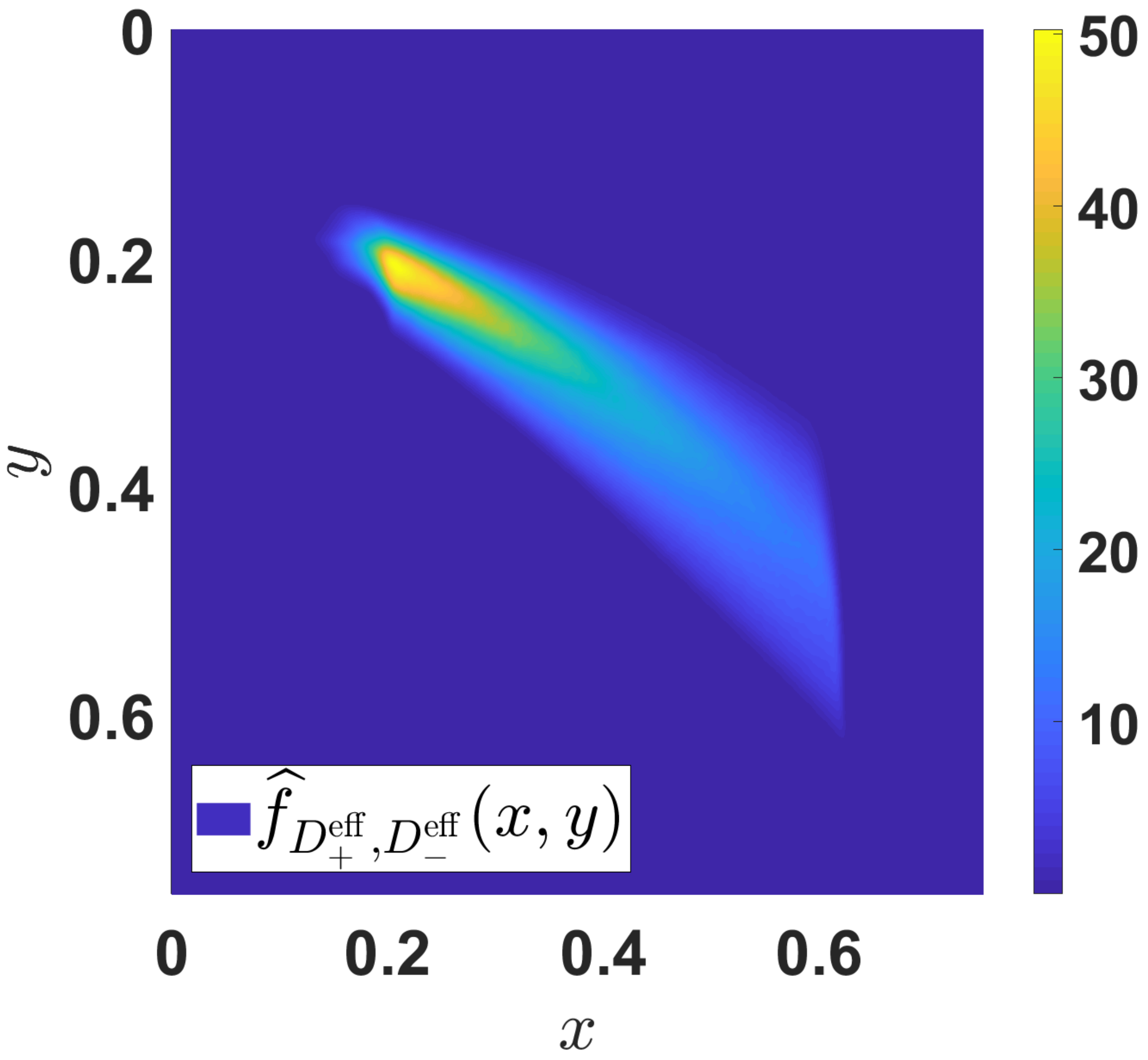}%
  \hfill
  \includegraphics[width=0.45\textwidth]{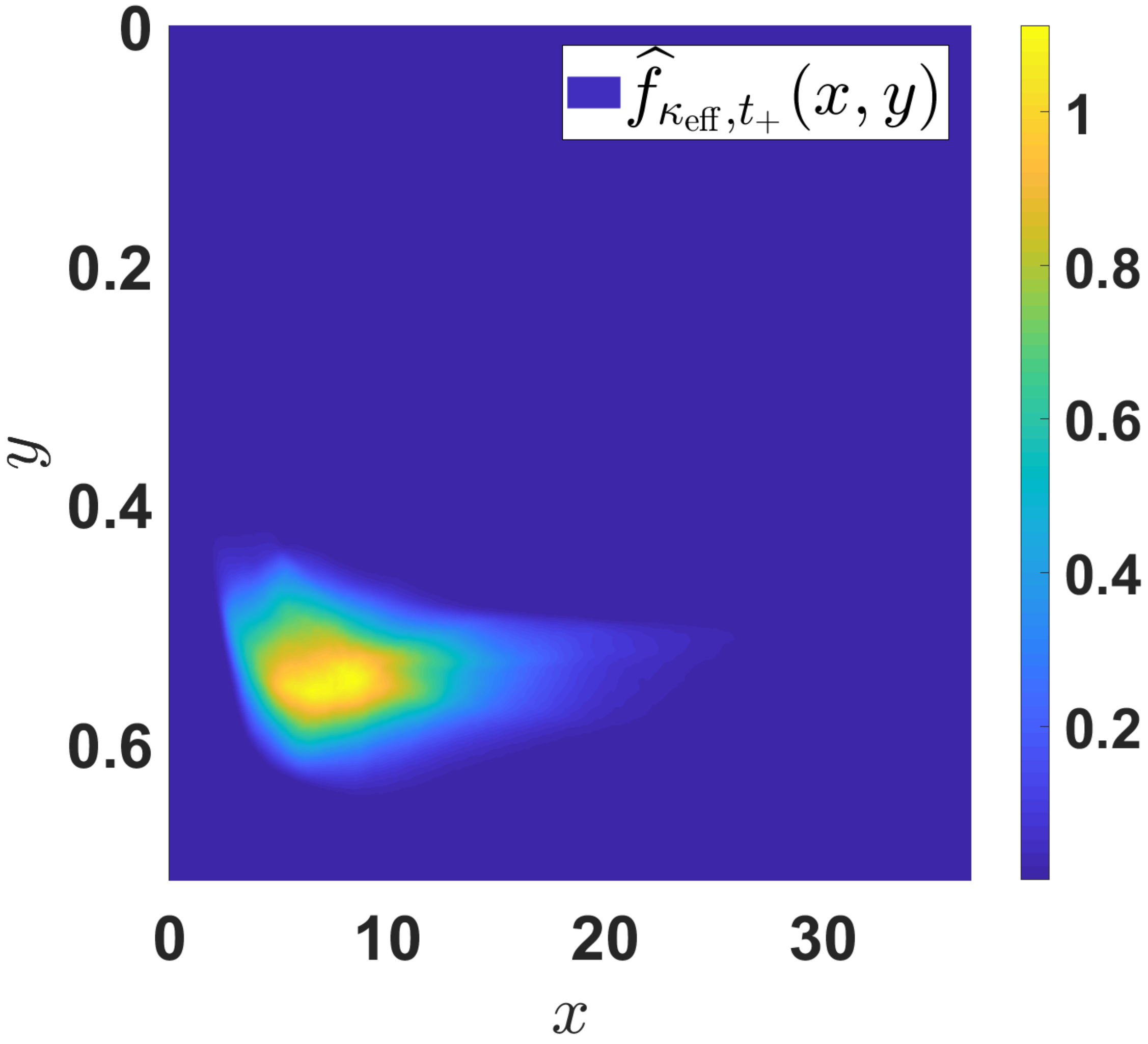}\\

  \caption{Estimated joint densities for all combinations of the QoI variables \cref{eq:qoi-vars}, based on $M=\num{1e7}$ samples of each respective output computed with the GINN, exhibit complex relationships that are non-Gaussian and skewed. The generation of faithful multivariate densities for uncertainty propagation requires surrogate modeling to produce a sufficient number of samples for the application of interest.}
  \label{fig:qois-joint-densities}
\end{figure}

Density estimates \cref{eq:kde_Gaussian_diag}, together with appropriate confidence intervals, provide a great deal of insight into the propagation of uncertainties that are lacking in summary statistics. This is especially true here because of the skewed and non-Gaussian nature of the QoIs \cref{fig:qois-marginal-densities_forward,fig:qois-joint-densities}. In \cref{fig:qois-marginal-densities_forward}, we plot a KDE for each QoI based on $M=\num{8e3}$ samples simulated using the BN PDE with the co-simulation forward model framework and on $M=\num{1e7}$ samples using the GINN. For each KDE, optimal kernel bandwidths are again chosen using the Improved Sheather--Jones method. The structured priors over broad ranges in \cref{tab:independent-params} produce a wide landscape of effective macroscopic dynamics observed in \cref{fig:qois-marginal-densities_forward,fig:qois-joint-densities}. A qualitative comparison of the densities in \cref{fig:qois-marginal-densities_forward} reveals that the GINN is making faithful predictions that do not include spurious features observed with cost-limited data obtained from the physics-based model simulation. Moreover, we can report the KDE with significantly increased confidence. That is, the confidence interval for the KDE based on the large number of samples enabled by the GINN is vanishingly small. In \cref{fig:qois-joint-densities}, we again observe the highly skewed and non-Gaussian nature of the joint densities for all combinations of QoIs based on $M=\num{1e7}$ samples from the GINN. Likewise (although it is not displayed in the plot), we report that the confidence intervals for each corresponding density are vanishingly small. As the surrogate model enables us to generate a large amount of io sample pairs, we do not require bootstrapping techniques to construct these KDEs or their corresponding confidence intervals.

\section{Conclusions and future work}
\label{sec:concl}

We developed a domain-aware surrogate model for simulation-based decision-making in complex multiscale and multiphysics systems that leverages well-known tools from deep learning and stochastic/predictive modeling. By embedding a probabilistic graphical model, specifically, a Bayesian network (BN), into a deterministic homogenized (i.e.,~physics-based) model for effective ion diffusion in an electrical double layer (EDL) supercapacitor, we distilled a Graph-Informed Neural Network (GINN) surrogate that incorporates domain knowledge via structured priors on tunable, possibly correlated, control variables (CVs) and bypasses computational bottlenecks in the physics-based model by replacing them with a NN. The complementary BN partial differential equation (BN PDE) model provided a context for interpreting the GINN's predictions.

Our analysis yielded the following major conclusions.
\begin{enumerate}
    \item Both the correlated CVs and output quantities of interest (QoIs) have non-Gaussian and skewed probability distributions, necessitating their characterization using probability density functions (PDFs) rather than via summary statistics.
    \item GINNs are able to leverage a ``sparse data investment'' to yield ``big data returns'': training on a small set of expensive-to-compute physics-based input-output (io) data, GINNs can cheaply generate a large set of output data to predict QoIs with confidence. This yields several orders of magnitude in computational cost savings compared to using physics-based models alone.
    \item Because GINNs can generate io data fast, they can be deployed to estimate the marginal and joint distributions of mutually correlated QoIs with tight confidence intervals.
\end{enumerate}

In a follow-up study, we plan to use the surrogate-based KDEs in \cref{fig:qois-marginal-densities_forward} and \cref{fig:qois-joint-densities} to perform data-driven UQ tasks including differential mutual information-based sensitivity studies according to the framework in \cite{UmHallEtAl:2019bn}, as opposed to variance-based approaches (e.g.,~\cite{UmZhangKatsoulakisEtAl:2017aa}), given the non-Gaussian nature of the QoIs. 

The weights and biases of the GINN were determined based on a single run of the entire algorithm including generation of training data with the BN PDE, fitting the GINN with the training set and evaluating its test (generalization) error on an independent test data set. Instead, we could perform, say, $N_\text{real}$ realizations of the overall procedure, each with a different (but same size) training set with inputs stemming from another set of CV realizations. Then a different set of GINN parameters would be learned, and testing the resulting network would yield a different value for the test loss for the same set of test data. This way, we could build up a PDF for the test loss, which would facilitate comparison of test set predictivity for different complexities of the GINN (e.g., using a different number of hidden layers or neurons per hidden layer), to find the GINN configuration that has the smallest $N_\text{real}$-averaged test loss. A similar reasoning was followed in \cite{Taverniers:2015ml} in the context of a PDE based surrogate for the two-dimensional nearest-neighbor kinetic Ising model.

\section{Acknowledgments}
A portion of this research was undertaken when E.~H. was a postdoctoral research scientist in the Chair of Mathematics for Uncertainty Quantification at RWTH Aachen University, Germany and was partially supported by the Alexander von Humboldt Foundation. The research of  M.~K. was partially supported by the HDR-TRIPODS program of the National Science Foundation (NSF) under grant CISE-1934846 and  by the Air Force Office of Scientific Research (AFOSR) under grant FA-9550-18-1-0214. The research of D.~T. was partially supported by by the Air Force Office of Scientific Research (AFOSR) under grant FA9550-18-1-0474 and by a gift from TOTAL.

\appendix  
  
\section{Physics-based model for supercapacitor dynamics}%
\label{sec:forward-model}%

EDL capacitors or EDLCs, as opposed to electrochemical pseudocapacitors, are supercapacitors that rely on the large capacitance of the EDL formed around the surface of their electrodes, typically comprised of a carbon-based hierarchical nanoporous material, for their energy storage capabilities \cite{Beguin:2013}. State models for EDLC cells governing the EDL behavior can be parametrized by effective constants obtained through rigorous mathematical homogenization \cite{ZhangTartakovsky:2017np} and related to the diffusion of ions in the cell's electrolyte. Rather than relying on phenomenological relations between microscopic parameters and their macroscopic counterparts, such homogenization or upscaling techniques \cite{ZhangEtAl:2015od,LingTartakovskyBattiato:2016dc} comprise an \emph{ab initio} approach for deriving macroscopic descriptors which allows them to establish the limits of applicability of the resulting macroscale model.

We consider a supercapacitor with electrodes consisting of a hierarchical nanoporous material $\mathcal{V}$ with characteristic length $L$ that is composed of a pore space $\mathcal{P}$ and impermeable (typically carbon) structure $\mathcal{S}$, i.e.,~$\mathcal{V} = \mathcal{P} \cup \mathcal{S}$ \cite{ZhangTartakovsky:2017np}. The pore space $\mathcal{P}$ is filled with an electrically neutral electrolyte with concentrations of positive ions or \emph{cations} (e.g.,~tetraethylamonium ions) and negative counter ions or \emph{anions} that evolve in both space and time. The characteristic length scale for this evolution, i.e.,~a typical pore diameter, is denoted by $\ell$ and satisfies $\epsilon \defeq \ell/L \ll 1$. The interaction of the electrolyte ions with static charges at the fluid-solid interface $\Gamma$ gives rise to an electrical double layer (EDL). A representative unit cell $\mathcal{U}$ consisting of pore space $\Upore$, structure $\Ustructure$, and fluid-solid interface $\Uinterface$ is illustrated in \cref{fig:pore-geom} adapted from \cite{ZhangTartakovsky:2017np}. We are interested in understanding macroscopic material properties that provide a homogenized (continuum-scale) description of the ion diffusion in the electrolyte, which was derived from a microscale formulation based on the Nernst--Planck equation in \cite{ZhangTartakovsky:2017np}. The latter describes the motion of the electrolyte ions under the influence of a concentration gradient and the electric field associated with the gradient of the electric potential in the EDL. The relevant QoIs are then the effective diffusion coefficients of the anions and cations in the electrolyte, and two derived quantities, the electrolyte conductivity and transference number.

\subsection{Effective ion diffusion coefficients}%
\label{sec:effective-diffusion}%

The effective (continuum-scale) diffusion coefficients of the cations and anions are second-order semipositive-definite tensors given by
\begin{equation}
  \label{eq:effective-diffusion} \Deffpm \defeq
  \frac{\mathcal{D}\omega}{G_{\pm}} \int_{\Upore} e^{\mp z \hphiEDL}
  (\eye + \nabla_{\mathbf{y}}\bchi_{\pm}^{\top}) \dd{\mathbf{y}}\,,
  \quad G_{\pm} \defeq \int_{\Upore} e^{\mp \hphiEDL} \dd{\mathbf{y}}.
\end{equation}
Here $\mathcal{D}$~[\si{\Length^2/\Time}]\footnote{We use \si{\Length} for dimensionless units of length and \si{\liter} for liters.} is the molecular diffusion coefficient of both ion species in the electrolyte;
\begin{equation}
  \label{eq:porosity}
  \omega \defeq \| \mathcal{P}\| / \| \mathcal{V} \|
\end{equation}
is the material porosity; $z$~[-] is the ion charge (valence);\footnote{For simplicity of presentation, we assume ion charge symmetry ($z_{+} = - z_{-} = z$) and equality of dissociation constants ($\nu_{+} = \nu_{-} = \nu$), i.e., the electrolyte salt is completely dissociated into cations and anions which have equal but opposite charges. The analysis can be easily extended to multicomponent and/or asymmetric electrolytes.} $\eye$ is the identity matrix; and $\bchi_{\pm}(\mathbf{y})$ are $\mathcal{U}$-periodic vector functions arising from the homogenization closure equations, i.e.,~that solve the boundary-value problems
\begin{subequations}
  \label{eq:closure}
  \begin{equation}
    \nabla_{\mathbf{y}} \bigl( e^{\mp z \hphiEDL}(\eye +
    \nabla_{\mathbf{y}} \bchi_{\pm}^{\top}) \bigr) =
    \boldsymbol{0}\,, \quad \mathbf{y} \in \Upore\,;    
  \end{equation}
  \begin{equation}
    \mathbf{n}(\eye + \nabla_{\mathbf{y}} \bchi_{\pm}^{\top}) =
    \boldsymbol{0}\,,
    \quad \mathbf{y} \in \Uinterface\,;
  \end{equation}
  \begin{equation}
    \int_{\Upore} \bchi_{\pm} \dd{\mathbf{y}} = \boldsymbol{0}\,.
  \end{equation}
\end{subequations}
In \cref{eq:closure}, $\hphiEDL$ is a non-dimensional formulation of the EDL potential $\varphi_{{\text{EDL}}}$~[\si{\volt}], that is,
\begin{equation*}
  \label{eq:phi-hat-edl}
  \hphiEDL = \frac{F \varphi_{\text{EDL}}}{RT}\,,
\end{equation*}
where $F = 96485$~\si{\coulomb\per\mol} is the Faraday constant, $R$~[\si{\joule\per\mol\per\kelvin}] is the gas constant, and $T$~[\si{\kelvin}] is the temperature. Under the assumption that the spatial variability of $\hphiEDL$ is confined to the nanoscale, it is found by solving
\begin{subequations}
  \label{eq:EDL-potential}
  \begin{equation}
    \hat{\nabla}^2 \hphiEDL = \frac{\ell^2
      \hat{c}_{\text{b}}}{\epsilon^2 \lambda_D^2} \sinh(z \hphiEDL)\,, \quad
    \mathbf{\hat{y}} \in \Upore\,; 
  \end{equation}
  \begin{equation}
    \hphiEDL =
    \hphiG\,, \quad \mathbf{\hat{y}} \in \Uinterface\,,
  \end{equation}
\end{subequations}
where $\ell$~[\si{\Length}] is the characteristic pore size; $\hat{c}_{\text{b}}\defeq c_\text{b}/\cin$~[-] with $c_\text{b}$~[\si{\mole\per\liter}] a characteristic ion concentration in the system (e.g., the initial or average concentration) and $\cin$~[\si{\mole\per\liter}] the initial ion concentration; the Dirichlet boundary condition
\begin{equation*}
  \hphiEDL = \hphiG = \frac{F \phiG}{RT}
\end{equation*}
arises from assuming that the surface $\Gamma$ carries a constant electric (zeta) potential $\phiG $ which is satisfied if the solid matrix is highly conductive as in the case of carbon aerogels (\cite{YingEtAl:2002ag}); and $\lambda_D$~[\si{\Length}] is the Debye length, a characteristic length of the EDL which in nanoporous materials is of the same order as the characteristic pore size $\ell$. The Debye length is given by
\begin{align}
  \label{eq:lambda_D}
  \lambda_D =
  \sqrt{\frac{RT\mathcal{E}}{2F^2 z^2 \nu \cin}}\,,
\end{align}
where $\mathcal{E}$~[\si{\farad\per\meter}] is the absolute permittivity of the solvent and $\nu$~[-] is the dissociation constant. We set $c_\text{b}=\cin$ (i.e.,~$\hat{c}_b = 1$) in our numerical experiments, and compute $\phiG$ by solving the transcendental equation
\begin{subequations}
  \label{eq:transc}
  \begin{align}
    \label{eq:transc:phiG}
    \phiG = \frac{V}{2} - \varphi_\text{ecm} -
    \frac{\sigma}{C_\text{H}},
  \end{align}%
  where $V$ is the external voltage, $\varphi_\text{ecm}$ is the electrocapillary maximum, $C_\text{H}$ is the Helmholtz capacitance, and $\sigma$ is the surface charge density given by
  \begin{equation}
    \label{eq:surf-charge}
    \begin{split}
      \sigma &= \sqrt{4\mathcal{E}
        RTI}\sqrt{\cosh\left(\frac{\mathrm{e}\phiG}{k_{\text{B}}T}\right)
        -
        \cosh\left(\frac{\mathrm{e}\varphi_\text{min}}{k_{\text{B}}T}\right)},
      \\ I &= z^2 C, \quad \varphi_\text{min} =
      \min_{\mathbf{y}\in\Upore} \varphi_\text{EDL}(\mathbf{y}).
    \end{split}
  \end{equation}
\end{subequations}
In \cref{eq:surf-charge}, $\mathrm{e}$~[\si{\coulomb}] is the elementary charge, $k_{\text{B}}$~[\si{\joule\per\kelvin}] is the Boltzmann constant, $I$ is the ionic strength, $C$ is the macroscopic ion concentration, and $\varphi_\text{min}$ is the midplane potential computed by solving \cref{eq:EDL-potential}. Since the latter requires knowledge of the boundary potential $\phiG$ which we are solving \cref{eq:transc} for, this would require an iterative procedure. Instead, we assume a value of 0.01 for $\varphi_{\text{min}}$. Additionally, for simplicity we set $C = \cin$, i.e.,~equal to the initial ion concentration. Using an initial guess of 0.3 for $\phiG$, performing a constrained ($\phiG\leq 0.5$) nonlinear optimization in \textsc{MATLAB}\textsuperscript {\textregistered} resulted in physically reasonable solutions for $\phiG$ over the ranges of $T$ and $\cin$ considered. The values of the various constants in \cref{eq:transc} are given in \cref{tab:consts}.

\begin{table}[htb]
  \centering
  \caption{Values for the constant parameters in \cref{eq:transc}. For simplicity we take these parameters as fixed values (cf.~\cref{eq:inputs-dependent:phiG} where these could also be incorporated as hyperparameters).}
  \label{tab:consts}
  \begin{tabular}{clSc}
    \toprule%
    \multicolumn{1}{l}{Variable} & {Meaning} & {Value} & {Units}\\%
    \midrule%
    $V$ & external voltage & 3 & \si{\volt}\\
    $C_\text{H}$  & Helmholtz capacitance & 0.45 &\si{\farad\per\m^2} \\
    $\mathcal{E}$  & solvent's dielectric constant & 6.9e-11 & - \\
    $\varphi_\text{ecm}$  & electrocapillary maximum & 0.1 &\si{\volt}\\
    $\varphi_\text{min}$  & midplane potential & 0.01 &\si{\volt} \\
    \bottomrule%
  \end{tabular}
\end{table}
  
\begin{figure}[htb] \centering
  \includegraphics[width=0.4\textwidth]{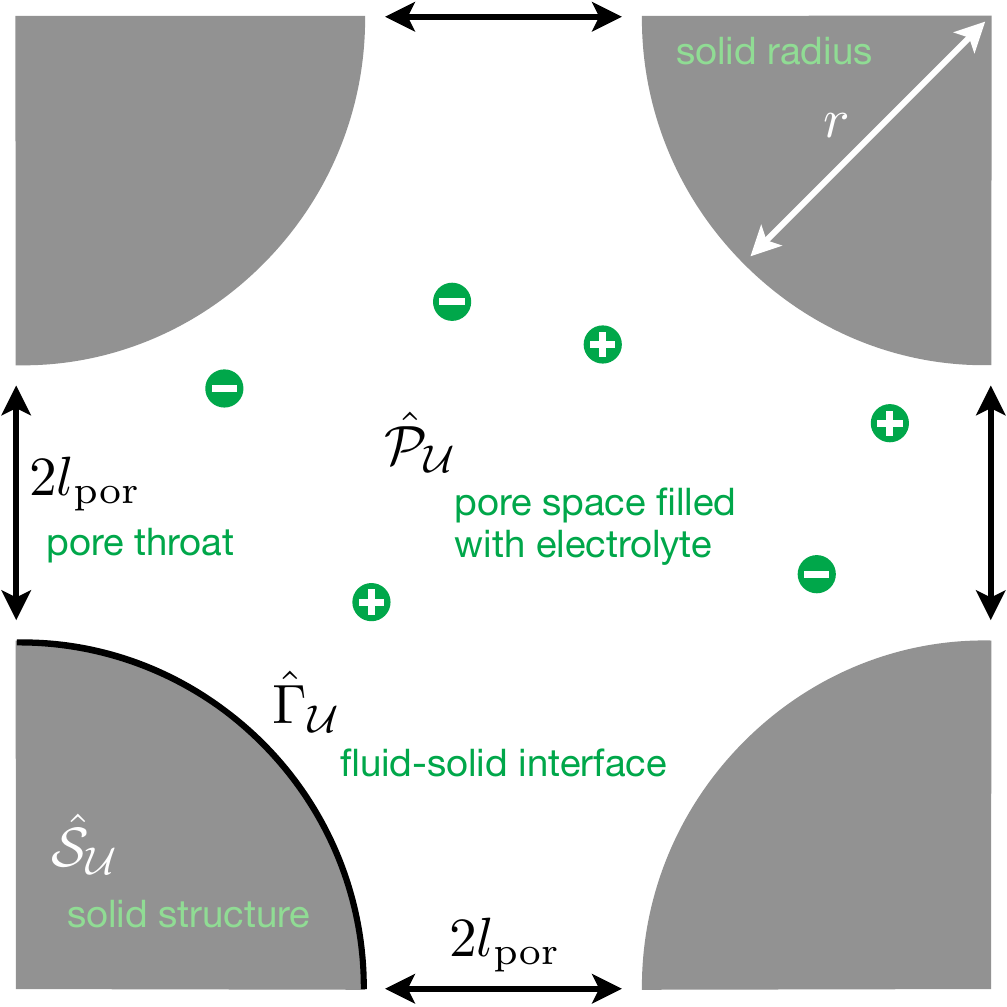}
  \caption{Unit cell $\mathcal{U}$ for a homogeneous isotropic nanoporous material containing circular impermeable structures $\Ustructure$ of radius $r$ separated by pore space $\Upore$ with pores of throat size $2 \lpor$ in both the transverse and longitudinal directions. The fluid-solid interface $\Uinterface$, involved in the formation of the EDL, is a key component in the investigation of EDLC cells. Data-driven UQ for simulation-aided design of nanoporous metamaterials is crucial to data-centric approaches for engineering advanced supercapacitors for long-term energy storage applications.}
  \label{fig:pore-geom}
\end{figure}

As opposed to the typical treatment that expresses effective diffusion through phenomenological relations, the rigorous derivation of the effective diffusion tensors in \cref{eq:effective-diffusion} using homogenization theory enables one to express them in terms of pore-scale geometry and processes including the EDL potential. In particular, we observe that only the diagonal elements of $\Deffpm$ are non-zero since off-diagonal elements of $\nabla_{\mathbf{y}} \bchi_{\pm}^\top$ are zeros in \cref{eq:closure}. Together with the fact that we consider a homogeneous isotropic nanoporous material (see unit cell in \cref{fig:pore-geom}), the diffusion coefficients in \cref{eq:effective-diffusion} become scalars,
\begin{equation}
  \label{eq:effective-diffusion-qois}
  \Deffp \qquad \text{and} \qquad \Deffm \,.
\end{equation}
The assumption of isotropy is reflected in the fact that the pore throat size is identical in both directions (\cref{fig:pore-geom}) and can be expressed in terms of the solid radius $r$ and porosity $\omega$ via
\begin{align}
  \label{eq:lpor}
  \lpor = -r + 0.5\sqrt{4r^2 + 4r^2\cdot
  \left[\frac{\pi}{4\cdot (1 - \omega)} - 1\right]},
\end{align}
where $\lpor$ refers to the \emph{half} pore throat size.\footnote{We take the \emph{half} instead of the \emph{full} pore throat size as an input parameter for reasons of convenience.}

\subsection{Effective electrolyte conductivity and transference
  number}%
\label{sec:effective-conductivity-transference}%

The effective diffusion coefficients in \cref{eq:effective-diffusion} are used to compute other key QoIs including the effective electrolyte conductivity $\keff$ [\si{\milli\siemens\per\cm}],
\begin{equation}
  \label{eq:kappa-eff}
  \keff \defeq \nu z^2 \frac{F^2 \cin}{RT}
  (\Deffp + \Deffm) \,,
\end{equation} and the transference number $\tp$ [-] (fraction of the
current carried by the cations),
\begin{equation}
  \label{eq:transference}
  \tp \defeq \frac{\Deffp}{\Deffp + \Deffm}\,.
\end{equation}

\subsection{Three-equation model governing the EDL}%
\label{sec:three_eqn_EDL}%

The effective ion diffusion parameters $\Deffp$, $\Deffm$, $\keff$, and $\tp$ serve as coefficients in a three-equation model (\cite{Newman:2012,VerbruggeLiu:2005ed,ZhangTartakovsky:2017np}) for macroscopic state variables that characterize the behavior of EDLC cells: electrolyte ionic concentration $C(x,t)$, electrolyte potential $\Phi(x,t)$, and electric potential of the solid phase $\Phi_s(x,t)$, satisfying,
\begin{subequations}
  \label{eq:edlc-state-model}
  \begin{equation}
  \mathcal{C}_{\text{EDL}} \frac{\partial(\Phi_s - \Phi)}{\partial t}
  = \frac{\partial}{\partial x} \left( \sigma_s \frac{\partial \Phi_s}{\partial x} \right)\,,
\end{equation}
\begin{equation}
  \frac{\partial}{\partial x} \left( \sigma_s \frac{\partial \Phi_s}{\partial x} + \keff \frac{\partial \Phi}{\partial x} + \keff RT \frac{2 \tp - 1}{z F} \frac{\partial \ln C}{\partial x} \right) = 0\,,
\end{equation}
\begin{equation}
  \omega \frac{\partial C}{\partial t} = \frac{\partial}{\partial x} \left( \frac{2 \Deffp \Deffm}{\Deffp + \Deffm} \frac{\partial C}{\partial x} - \alpha \frac{\partial(\Phi_s - \Phi)}{\partial t}\right)\,,
\end{equation}
\end{subequations}
for $x \in [0,L]$ and $t > \infty$, subject to suitable boundary conditions where $\sigma_s$ is the electric conductivity of the solid phase, $\mathcal{C}_{\text{EDL}}$ is the EDL capacitance, and $\alpha = \alpha(x)$ is a piecewise defined function of $\tp$ that varies according to electrode thickness and separator thickness. In the remainder of this work, we will use their normalized counterparts $\Deffp/\mathcal{D}$ and $\Deffm/\mathcal{D}$, but retain the notation $\Deffp$ and $\Deffm$ to refer to these normalized effective diffusion coefficients.

\Cref{eq:effective-diffusion,eq:porosity,eq:closure,eq:EDL-potential,eq:lambda_D,eq:transc,eq:effective-diffusion-qois,eq:lpor} and \cref{eq:edlc-state-model} constitute two stages of a complex model for supercapacitor dynamics in EDLC cells that includes nonlinear multiscale and multiphysics interactions. For constructing our BN PDE model, we focus only on the first stage covered in \cref{sec:effective-diffusion} and \cref{sec:effective-conductivity-transference}. A similar reasoning can be followed for the EDLC state model detailed in the current section.

\bibliography{ginns}%

\end{document}